\documentclass[aps,prb,twocolumn,floatfix]{revtex4-1}
\usepackage{graphicx,epsfig,array,color}
\usepackage{color,graphicx,amssymb,amsmath}
\begin{document}

\title{Testing the Monte Carlo - Mean Field approximation in the one-band Hubbard model}

\author{Anamitra Mukherjee$^1$}
\author{Niravkumar D. Patel$^1$}
\author{Shuai Dong$^2$}
\author{Steve Johnston$^1$}
\author{Adriana Moreo$^{1,3}$}
\author{Elbio Dagotto$^{1,3}$}

\affiliation{$^1$Department of Physics and Astronomy, The University of Tennessee, Knoxville, Tennessee 37996, USA}
\affiliation{$^2$Department of Physics, Southeast University, Nanjing 211189, China}
\affiliation{$^3$Materials Science and Technology Division, Oak Ridge National Laboratory, Oak Ridge, Tennessee 37831, USA}

\date{\today} 

\begin{abstract}  
The canonical one-band Hubbard model is studied using a computational method that mixes the Monte Carlo procedure with the mean field approximation. This technique allows us to incorporate thermal fluctuations and the development of short-range magnetic order above ordering temperatures, contrary to the crude finite-temperature Hartree-Fock approximation, which incorrectly predicts a N\'eel temperature $T_N$ that grows linearly with the Hubbard $U/t$. The effective model studied here contains quantum and classical degrees of freedom. It thus belongs to the ``spin fermion'' model family widely employed in other contexts. Using exact diagonalization, supplemented by the traveling cluster approximation, for the fermionic sector, and classical Monte Carlo for the classical fields, the Hubbard $U/t$ vs. temperature $T/t$ phase diagram is studied employing large three and two dimensional clusters.  We demonstrate that the method is capable 
of capturing the formation of local moments in the normal state 
without long-range order, the non-monotonicity of $T_N$ with increasing $U/t$, the development of gaps and pseudogaps in the density of states, and the two-peak structure in the specific heat. Extensive comparisons with determinant quantum Monte Carlo results suggest that the present approach is qualitatively, and often quantitatively, accurate, 
particularly at intermediate and high temperatures. Finally, we study the Hubbard model including plaquette diagonal hopping (i.e. the $t-t^\prime$ Hubbard model) in two dimensions and show that our approach allows us to study low temperature properties where determinant quantum Monte Carlo fails due to the fermion sign problem.
Future applications of this method include multi-orbital Hubbard models such as those needed for iron-based superconductors.
\end{abstract}
\maketitle

\section{Introduction}
 The study of strongly correlated electrons continues attracting the interest of the condensed matter community.\cite{tokura,complex} Theoretical studies in this area of research mainly use model Hamiltonians since there are no {\it ab-initio} techniques that
 can handle with sufficient accuracy the correlation effects caused by the Coulombic charge repulsion among the electrons.  The case of the Hubbard model with only one active orbital ($d_{x^2-y^2}$) has been 
 widely studied in the context of copper-based  high temperature superconductors, and a variety of interesting results and predictions have  been unveiled.\cite{RMP94,scalapino,lee} A large fraction of those studies, however, arise from approximate analytic many-body techniques that are difficult  to control since there is no obvious small parameter to guide expansions when one is dealing with correlated electrons. 
 For this reason, considerable efforts have been devoted to the use of computational techniques  to study Hubbard-like models.\cite{RMP94} 
 Alas, these computational methods are not without severe limitations as well. For example, the  Lanczos method is restricted to small clusters\cite{RMP94} while the density matrix renormalization group (DMRG)  is restricted to quasi one-dimensional systems.\cite{dmrg} An alternative is the determinant quantum Monte Carlo (DQMC) technique, \cite{dqmc,WhitePRB1989,paiva10} which can handle the one-orbital
 Hubbard model in dimensions larger than one and employing clusters of a reasonable size.
 This technique has been applied in numerous occasions, leading recently also to studies in the context of optical lattices.\cite{optical-lattices,paiva,kozik} DQMC presents the infamous ``sign problem'', however, which severely restricts its range of  applicability. For instance, deviations from the particle-hole symmetric model, such as when electronic hopping beyond nearest-neighbors are introduced, or when doping away from half-filling is attempted, severely restricts the temperature range where DQMC can be applied.\cite{sign,WhitePRB1989}

 The limitations of our computational arsenal to deal with Hubbard-like
 models have been exposed even more dramatically by the recent discovery of the iron-based 
 high temperature
 superconductors.\cite{johnston,stewart,natphys12,RMP13} 
 While considerable theoretical progress has been achieved in this context via the use
 of mean field approximations of several varieties,\cite{johnston,stewart,RMP13,peter,si} 
 computational work has been severely 
 restricted. This is mainly because of the need to incorporate several $3d$ iron 
 orbitals in the model Hamiltonian. It is well known that the Hubbard model 
 for pnictides must have a minimum of two iron orbitals: $d_{xz}$ and $d_{yz}$, while most experts 
 agree that at least a third orbital $d_{xy}$
 should also be incorporated.\cite{three}
 Moreover, the crystal structure indicates that hopping amplitudes must
 involve both Fe-Fe nearest and next-nearest neighbors processes. All these factors are detrimental
 to the performance of Lanczos, DMRG, and DQMC techniques, and the applications of these
 methods have been limited in the context of the iron-based superconductors. In fact, in a recent
 review,\cite{natphys12} a crude drawing of the phase diagram of a multi-orbital Hubbard model was
 sketched ``by hand'' based on physical expectations, but this prediction has 
 yet to be confirmed due to the lack of reliable techniques for the calculations. 
 
 Motivated by the above mentioned difficulties in handling the full problem, 
 simplified versions of multi-orbital Hubbard models have been
 recently used in a number of contexts. For instance, in the colossal 
 magnetoresistive manganites\cite{physrep,salamon} the Double Exchange (DE)
 model separates the five $3d$ orbitals of Mn into mobile and localized degrees
 of freedom.\cite{physrep} This is compatible with the splittings caused
 by the crystal field; thus the separation of mobile and localized carriers is natural. 
 Moreover,
 it has been shown that the localized spins, related to the $t_{2g}$ orbitals, 
 can be approximated accurately by a classical spin.\cite{physrep,dmrgmanga} 
 Extensive studies employing computational Monte Carlo techniques have provided 
 ample evidence that this type of models can properly capture the physics 
 of manganites.\cite{yunoki,science99,kumar,dong1,dong2,liang11,sen} 
 In contrast, employing a full five-orbital Hubbard model would have been 
 impractical for the manganese oxides.
 
 The DE model is a well-known example of a more general family of models referred
 to as ``spin fermion'' models, where ``spin'' denotes the localized degree of freedom
 and ``fermion'' denotes the mobile one. As in the DE case, the localized spin
 is considered classically in practice to allow for reasonable computational studies. 
 Historically, the success of the DE model treated 
 computationally has inspired the use of spin-fermion models 
 for the cuprates as well. Spin fermion models for Cu oxides are technically 
 similar to DE models and they have been
 able to reproduce features of the one-orbital Hubbard model, such
 as the dominance of $d$-wave pairing tendencies away
 from half-filling.\cite{buhler1,buhler2,mora1,mora2,mora3} 
 A similar approach has been used in the context of the Bogoliubov de Gennes (BdG) equations,
 allowing for  the study of regimes beyond weak coupling BCS.\cite{BdG0,BdG1,BdG2,BdG3} 
 It should be stressed that none of the spin-fermion models, either in the manganite or cuprate
 context, exhibit a sign problem. Thus, computational studies are possible 
 at any electronic density, temperature, and range of electronic hopping. 
Moreover,
 in spin fermion models dynamical observables can be easily obtained, contrary to dynamical observables
 in full Hubbard models that require calculations in imaginary time 
and a subsequent transformation to real frequency.\cite{MaxEnt}
 
 Spin fermion models seem to capture the qualitative essence of Hubbard models. 
 Typically, however, they are defined ``by hand'' in cases where the mobile-localized
 separation is intuitively expected but it is unclear how this separation truly occurs
 in practice. (This is contrary to the DE model, where $e_g$ and $t_{2g}$ 
 orbitals clearly separate the mobile electrons from the localized electrons). 
 Thus, a method for 
 constructing spin-fermion models systematically from their parent Hubbard models is desirable.
 This also would reveal the relationship between the effective couplings in the 
 spin-fermion models and those of the more fundamental Hubbard interactions 
 such as the repulsion $U$.
 
 In this publication we explore these issues in depth in the context of the
 repulsive one-orbital Hubbard model. The essence of the computational method 
 described here is to setup
 the mean field equations for the problem at hand, and then
 raise the mean field parameters, such as the effective staggered magnetic
 field that appears for an antiferromagnetic (AFM) state, to the level of 
 a classical variable, which is then treated via Monte Carlo 
 simulations at finite temperatures. These classical variables play the
 role of the ``spin'' in the resulting spin-fermion-like model. For 
 the ``fermions'' the resulting Hamiltonian is quadratic and can be solved
 numerically via library subroutines or other procedures. 
 To our knowledge, the first time that this methodology was proposed was 
 in a study of the competition between AFM and
 superconducting (SC) tendencies in the one-orbital 
 Hubbard model.\cite{BdG2} In this earlier work, both the staggered AFM field and 
 a complex field representing the SC order parameter deduced from the BdG equations 
 were introduced and handled via Monte Carlo simulations. Similar studies involving 
 competing AFM and SC states within the Heisenberg model were presented in 
 Refs.~\onlinecite{BdG1} and \onlinecite{BdG3}. 
 In more recent efforts, these 
 main ideas were also independently derived in detailed studies 
 of the Hubbard model on an anisotropic triangular lattice\cite{triangular} and
 on a geometrically frustrated face centered cubic lattice.\cite{fcc}
 Applications of the same approach but for the case of an attractive
 Hubbard interaction (negative $U$)   
 that leads to pairing and superconductivity have been reported 
 in Refs.~\onlinecite{BdG0,BdG1,BdG2,BdG3,tarat1,tarat2,tarat3}.
  
 As we will show, an interesting result is that this computational procedure
 captures the highly non-trivial non-monotonic behavior of  the 
 N\'eel temperature $T_N$ with increasing Hubbard $U$ at half-filling, in excellent
 agreement with DQMC. This is a dramatic
 improvement over standard Hartree-Fock mean field techniques that
 incorrectly predict a smooth increase of $T_N$ with $U$. This
 ``up and down'' behavior of $T_N$ with $U$ was also observed recently in a similar
 study of the negative $U$ Hubbard model~\cite{tarat2} (note that there is a mapping
 between positive and negative $U$), and in early studies
 of models for $d$-wave superconductivity with increasing pairing
 attraction.\cite{BdG0,BdG1,BdG2} In addition, 
 many other observables calculated within this approach, such as the specific heat,  
 are in qualitative, and often quantitative, agreement with DQMC, as shown below.
 Moreover, the spin-fermion model also allows for the calculation of 
 dynamical observables directly in real time and frequency.  We 
 demonstrate this here by calculating the single-particle density of states. Finally, 
 we further examine the utility of this approach 
 by examining the Hubbard model with longer range hopping. In this case, DQMC 
 cannot be applied due to a severe fermion sign problem but the
 new method is successful. 

 In summary, the simple combination
 of Monte Carlo and mean field methods
 allows for a proper treatment of the temperature effects in Hubbard
 models, including the study of regimes where the relevant correlations, such
 as the spin correlations, are of short range in space.
 Although it will be computationally demanding, after the success of the test presented
 here and in the other publications cited above, the method will be ready to be implemented
 for multi-orbital Hubbard models of relevance in, e.g., iron superconductors,  
 where virtually nothing is known about their thermodynamic behavior. 
 
 This paper is organized as follows. The Hubbard model and the technique  
 are discussed in Sec. II. 
 The technique is formally introduced by using the Hubbard-Stratonovich variables first employed in Ref.~\onlinecite{triangular}.  The main results are presented in Sec. III, starting with
 the case of three dimensions and its comparison with DQMC. 
 This is followed by a presentation of results for the two dimensional case, 
 as well as results for a Hubbard model with hopping beyond nearest neighbors 
 where DQMC suffers from a severe sign problem. We conclude in 
 Sec. IV with a brief summary and outlook.
 \vskip -2cm
\section{Model $\&$ method}
Let us start the specific application  of the ideas outlined in the Introduction 
by considering the one-band Hubbard model defined below (in a standard notation):
\begin{eqnarray}
H=H_o+H_1=-t\sum_{\langle i,j \rangle,\sigma}c^{\dagger}_{i,\sigma} c^{\phantom{\dagger}}_{j,\sigma}  +
U\sum_i n_{i,\uparrow}n_{i,\downarrow} 
\label{1}
\end{eqnarray}
To setup the formalism, it is convenient to perform 
a rotationally invariant decoupling of the interaction term in the following manner:\cite{method-4}
\begin{eqnarray} 
  n_{i,\uparrow}n_{i,\downarrow} &=& \frac{1}{4}(n_i^2)-S_{iz}^2 \nonumber\\
  &=& \frac{1}{4}(n_i^2)-({\bold S}_i \cdot \hat{\Omega}_i)^2.
\label{2}
\end{eqnarray}
Here, the spin operator is ${\bold S_i}=\frac{\hbar}{2}\sum_{\alpha,\beta} c^{\dagger}_{i,\alpha}
{\bold \sigma}^{\phantom\dagger}_{\alpha,\beta}c^{\phantom\dagger}_{i,\beta}$, $\hbar=1$,  
$\{\sigma^x,\sigma^y,\sigma^z \}$ are the Pauli matrices, and $\hat{\Omega}$ is an 
arbitrary unit vector. In the previous identity, we have used 
the fact that 
$({\bold S}_i\cdot \hat{\Omega}_i)^2=(S_{i,x})^2=(S_{i,y})^2=(S_{i,z})^2$.  
The expression in the last line of Eq. \ref{2} is rotationally invariant since it is in terms of the scalars $n_i$ and the dot product 
between ${\bold S}_i$ and $\hat{\Omega}_i$. 
It should be noted that there are other possible decouplings, but the formula above is the only one whose saddle point leads to the correct Hartree-Fock equations after implementing a Hubbard-Stratonovich (HS) decomposition. Below we will use the notation followed in recent literature.\cite{santos}. For the HS decomposition, let us start with the partition function $Z=Tre^{-\beta H}$. Here the trace is over all particle numbers and site occupations. $\beta=1/T$, with $k_B$ set to 1. We now divide the interval $[0,\beta]$ into $M$ equally spaced slices, defined by $\beta=M\Delta \tau$, separated by $\Delta \tau$ and labeled from 1 to $M$. For large M, $\Delta \tau$ is a small parameter and allows us to employ the Suzuki-Trotter decomposition, so that we can write $e^{-\beta(H_o+H_1)}=(e^{-\Delta \tau H_o}e^{-\Delta \tau H_1})^M$ to first order in $\Delta \tau$. 
Then using Eq.(2) and the Hubbard-Stratonovich identity, $e^{-\Delta \tau U \sum_i [\frac{1}{4}(n_i^2)-({\bold S}_i \cdot \hat{\Omega}_i)^2]}$, for a generic time slice $^\prime l^\prime$, can be shown to be proportional to,
\begin{eqnarray} 
\int {d\phi_i(l) d\Delta_i(l)
d^2\Omega_i(l)}\times \hskip 4.4cm& \nonumber\\
e^{-\Delta \tau [\sum_i(\frac{\phi_i(l)^2}{U}+i\phi_i(l)n_i+\frac{{\Delta_i(l)}^2}{U}
-2 {\Delta_i(l)}\hat{\Omega}_i(l).{\bold S_i})]}&\nonumber
\end{eqnarray}
Here we have introduced two auxiliary fields, $\phi_i(l)$ which couples to the charge density, and $\Delta_i(l)$ that couples to the spin density. We note that the integration over unit vector, $\hat{\Omega}_i(l)$ at every site shows the SU(2) invariance explicitly. We further combine the product $\Delta_i(l)\hat{\Omega}_i(l)$ into a new vector auxiliary field, ${\bold m_i}(l)$ at every site. This nomenclature is used from now on. Using the above decoupling for the quartic term in the expression of the partition function, we can write:
\begin{widetext}
\begin{equation}
Z= const. \times Tr \prod^1_{l=M} \int {d\phi_i(l) d^3m_i(l)}e^{-\Delta \tau [-t\sum_{\langle i,j \rangle,\sigma}c^{\dagger}_{i,\sigma} c^{\phantom{\dagger}}_{j,\sigma}+\sum_i(\frac{\phi_i(l)^2}{U}+i\phi_i(l)n_i+\frac{{\bold m_i(l)}^2}{U}
-2 {\bold m_i(l)}.{\bold S_i})]}
\end{equation}
\end{widetext}
In the above, trace `Tr' is over all particle numbers and site occupations as before . The continuous integrals are over the auxiliary fields, $\{\phi_i(l),{\bold m_i(l)}\}$ at every site and the argument $l$ denotes imaginary time slice label. The product over $l$ from M to 1 implies time ordered products over time slices, with the earlier times appearing to the right. Finally, the $d^3m_i(l)$ in the integral, implies integration over the amplitude and orientation of vector auxiliary fields, ${\bold m_i(l)}$.

This allows us to identify an effective Hamiltonian $H_{eff}$ in which fermions couple to auxialiary fields fluctuaing in both space and (imaginary) time. Typically this is the starting point of Quantum Monte Carlo (QMC) approaches. However for reasons discussed in the introduction, we take a different route by  making the following approximations. (i) We drop the $\tau$ dependence of the HS auxiliary fields and (ii) we use the saddle point value $i\phi_i=\frac{U}{2} \langle n_i\rangle$. This allows us to extract the following effective Hamiltonian ($H_{eff}$) where the fermions couple to the ``static'' HS field ${\bold m_i}$ and to the average local charge density:
\begin{eqnarray} 
 H_{eff}&=&H_o- \mu\sum_i 
n_i+\sum_i\frac{U}{2} \langle n_i \rangle n_i-\sum_i{\bold m_i}.{\bold \sigma_i}\nonumber\\
&+&\frac{1}{U}\sum_i{\bold m_i}^2-\frac{U}{4}\sum_i \langle n_i\rangle^2.
\end{eqnarray}
Here, $H_o$ contains the fermionic kinetic energy.  
The redefinition ${\bold m}_i\rightarrow\frac{U}{2}{\bold m}_i$ allows us to arrive 
to the final form of the effective Hamiltonian:

\begin{eqnarray} 
 H_{eff}&=&
H_o+\frac{U}{2} \sum_i(\langle n_i\rangle n_i-{\bold m_i}.{\bold \sigma_i})\\
&+&\frac{U}{4}\sum_i({\bold m_i}^2-\langle n_i \rangle^2)-\mu\sum_i n_i,
\nonumber
\end{eqnarray}
which is our effective model belonging to the spin-fermion family.

It should be noted that $H_{eff}$ coincides with the mean-field Hamiltonian at $T=0$, where ${\bold m_i}$ has the interpretation of the local magnetization. As discussed in the Introduction, to study the model at finite temperature, we simulate $H_{eff}$ by sampling the ${\bold m_i}$ fields via a classical Monte Carlo (MC) procedure.\cite{BdG2,tarat1,triangular} The main result of the present effort will arise when these MC results at finite temperature 
are compared against DQMC results. It will be demonstrated that retaining thermal fluctuations in the fields ${\bold m_i}$ leads to results well beyond simple Hartree-Fock mean field calculations at finite temperature $T$ and, more importantly, in good qualitative and sometimes quantitative agreement with DQMC.

While the HS fields are treated via MC methods, the quadratic fermionic sector still needs to be handled numerically.
The simplest and most widely employed method, starting with efforts in the manganite community to study  the double
exchange model,\cite{yunoki} is simply to carry out
an exact diagonalization (ED) of the fermions 
in a fixed classical ${\bold m_i}$ background, employing library subroutines. 
The ${\bold m_i}$ variables are then updated with a standard classical MC procedure 
where updates are accepted/rejected using the 
Metropolis algorithm. At a fixed temperature, this process is repeated until a thermalized regime is reached where
observables can be measured. 

Similarly as with the majority of techniques dealing with strongly correlated electrons, here there is no small parameter controlling the approximation. In particular there is no rigorous proof of convergence or bounded errors. However, as long as the mean field approximation employed as the starting point of the approximation (in the present example the Hartree Fock method) captures the essence of the ground state, then it is reasonable to assume that the present method will treat correctly the thermal fluctuations and associated short-range order tendencies above the critical temperature.

In the present work, we consider the case of half filling, 
where the total density is fixed by adjusting the chemical potential $\mu$.  
The Hamiltonian is studied on cubic lattices with 4$^3$ to 16$^3$ sites 
and with periodic boundary conditions. 
The magnetic
structure factors $S({\bf q})$ are used to perform finite lattice size scaling to extract
thermodynamic N\'eel temperatures in 3D.
Results obtained on two dimensional clusters will also be shown. 
All parameters are specified in units of the hopping 
$t$. In practice a total of 4000 MC system sweeps were 
performed: 2000 were used to thermalize the system, 
while the rest were used for calculating observables. 
A MC system sweep consists of sequentially visiting every 
lattice site  and updating the local ${\bold m_i}$ 
vector followed by the fermionic ED, and then 
accepting/rejecting the proposed local 
field change following the Metropolis algorithm. 
The local density $\langle n_i \rangle$ is computed from the 
eigenvectors after each diagonalization. 
In our calculation, 
we start the simulation at high temperature with a random 
configuration of ${\bold m_i}$ variables  
and then cool down to lower temperatures.  
To study the formation of local moments, as explained below, 
we start the MC runs at $T/t=100$ and cool down in steps of $\Delta T/t=1$ up to $T/t=1$. From $T/t=1$  
to 0.1, we use a step size of $0.1t$. Below this temperature, 
specifically from $0.1t$ to $0.005t$, we reduce further the
interval and use $\Delta T/t=0.05$. 
This slow process allows us to avoid metastable states or obtaining 
results that depend substantially on the initial conditions of 
the calculation.

To characterize the Hubbard model, a number of observables are computed during the MC procedure. 
In particular, we calculate the density of states (DOS), 
$ N(\omega) = \sum_{m} \delta(\omega-\omega_m)$, 
where $\omega_m$ are the eigenvalues of the fermionic sector and the summation runs up to $2 N^3$, 
i.e. the total number of eigenvalues of a $N^3$ system with spin. $ N(\omega)$ is calculated by implementing
the usual Lorentzian representation of the $\delta$ function. The broadening needed to obtain $N(\omega)$ 
from the Lorentzians is $\sim BW/2N^2$, where $BW$ 
is the fermionic bandwidth at $U=0$. 
Numerically for the 4$^3$ system, the broadening is about $0.09t$. Two hundred $N(\omega)$ samples 
are obtained from the 2000 measurement system sweeps at every temperature. We discard 10 MC steps between 
measurements to reduce self-correlations in the data. The 200 $N(\omega)$ samples are used to obtain 
the thermally averaged $\langle N(\omega)\rangle_T$ at a fixed temperature. These are further averaged 
over data obtained from 10-20 independent runs with different random number seeds.

Information regarding the N\'eel AFM order expected at half-filling 
is obtained from the magnetic structure factor,
\begin{equation}
S({\bold q}) = \frac{1}{N^2} \displaystyle\sum\limits_{i,j} e^{i {\bold q} \cdot ({\bold r}_{i}-{\bold r}_{j} )} 
\langle {{\bold S_i} \cdot {\bold S_j}}\rangle, 
\end{equation} 
where ${\bold q}=\{\pi,\pi,\pi\}$ is the wavevector of interest. 
The spins  ${\bold S_i}$ are constructed from the eigenvectors of the equilibrated configurations.

We also calculate the real space correlation function between the ${\bold S_i}$ vectors. 
This correlation function is defined as, 
\begin{equation}
C(|{\bold r}|)=\frac{1}{P}\sum\limits_{|{\bold r}|=|{\bold i}-{\bold j}|,a} (-1)^{|{\bold i}-{\bold j}|} \langle S^a_{i}S^a_{j}\rangle. 
\end{equation}
In $C(|{\bold r}|)$  
the summation runs over all P pairs of sites at a distance $|{\bold r}|$
and is normalized accordingly. The sum over $a$ runs over the three directions
$x$, $y$, and $z$.
\begin{figure}[t]
\centering{
\includegraphics[width=6.cm, height=6.cm, clip=true]{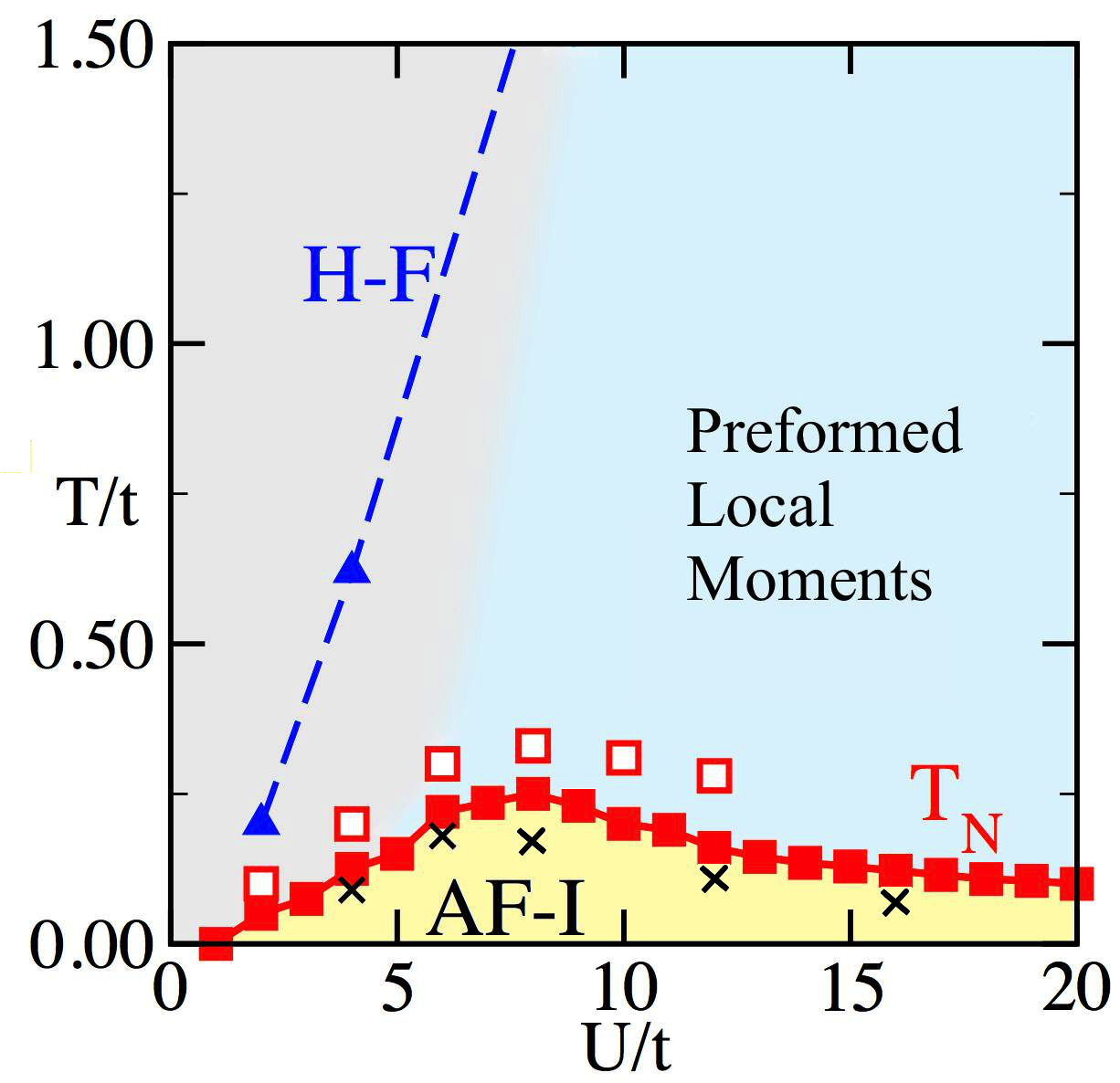}}
\caption{(color online)  The $T/t-U/t$ phase diagram for the one-band Hubbard model. 
The solid red squares show the dependence of $T_N$ on $U/t$ obtained using the MCMF technique 
on $4^3$ clusters. The crosses are estimations of $T_N$ 
obtained from finite-size scaling. The AF-I region denotes the N\'eel type AFM 
phase with long-range order and insulating characteristics.  
The open squares are the $T_N$ obtained from the DQMC method, 
from  Ref.~\onlinecite{muramatsu-1}.  
The light blue region depicts the regime of preformed local moments above the AF-I phase. 
The dashed line shows the $T_N$ obtained from the simplistic Hartree-Fock calculation 
at finite temperature where the critical temperature incorrectly grows linearly with $U/t$ 
at large $U/t$. The determination of the crossover between the gray and blue 
regions, and the fact that the MCMF local moment region coincides with HF $T_N$ 
at temperatures much larger than typical $T_N$ scales, are discussed in the text.}
\vspace{-0.0cm}
\label{rf-1}
\end{figure}
\begin{figure}[t]
\centering{
\includegraphics[width=9cm, height=10cm, clip=true]{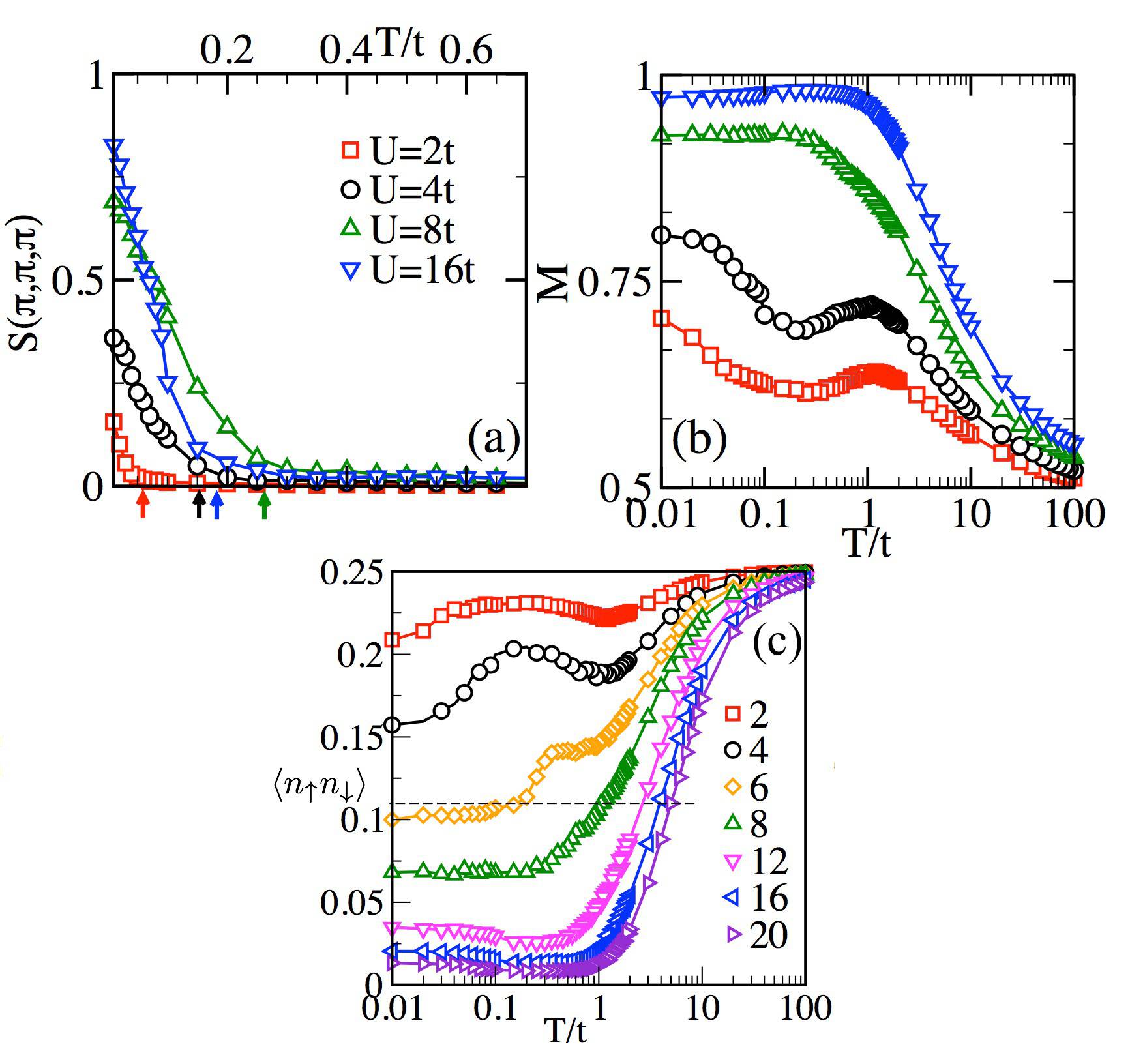}}
\caption{(color online)  (a) The magnetic structure factor $S({\bf q})$ 
for ${\bf q}=(\pi,\pi,\pi )$, obtained at various $U/t$'s as indicated. 
The data is from 4$^3$ clusters with 4000 MC sweeps at every temperature, 
while cooling the system down from high to low temperatures, as described 
in Sec. II. (b) The corresponding local moments, $M$, vs. temperature. 
We capture the feature that at large $U/t$ the peak in the moment size shifts 
to nonzero temperature. This effect, due to the setting of long range order, 
was reported before in the DQMC studies of Ref.~\onlinecite{scalettar-1}. 
At small $U$ the overall shape is also in good agreement with the DQMC data, 
indicating that the MCMF method indeed captures the essence of the problem. 
(c) shows the expectation value of double occupation for the various $U/t$'s 
indicated. The thin dashed line indicates a cutoff discussed in the text. }
\vspace{-0.0cm}
\label{rf-2}
\end{figure}
The distribution of the magnitude of $|{\bold m_i}|$ on the lattice is measured 
by the distribution function $P_q(|{m}|)$. This is defined as $ P_q(|{m}|)= \sum_{i} \delta(|m|-|{\bold m_i}|)$.  
For computational purposes, a Lorentzian representation with suitable broadening is used.  
$S({\bold q})$, $ P_q(|{m}|)$, and $ C(|{\bold r}|)$ are also averaged in the same manner 
as described for $N(\omega)$. We also compute the specific heat $C_v(U,T)=\frac{dE(U,T)}{dT}$ 
by numerically differentiating the average energy with respect to temperature. Other observables that
we measured are presented below.

\section{Results}

\subsection{The Three Dimensional Lattice}
Let us start the analysis of results with the $T/t-U/t$ 
phase diagram at half filling in 3D. 
In Fig.~\ref{rf-1}, we show the N\'eel temperature $T_N$ (solid squares; $4^3$ cluster) 
at different $U/t$'s obtained 
using the above described Monte Carlo - Mean Field (MCMF) method. The open squares are 
DQMC results, obtained from Ref. \onlinecite{muramatsu-1}. 
The most important characteristic of the MCMF results is that they 
correctly capture the ``up and down'' {\it non}-monotonic behavior of $T_N$ with increasing $U$. 
In particular, comparing our results against the standard Hartree-Fock (HF) mean field 
theory predictions for $T_N$ (shown in dashed blue) highlights the crucial role of thermal fluctuations. 
These fluctuations break down the uniform mean-field order with varying degree of ease as $U/t$ is 
changed. The MCMF method includes thermal 
fluctuations and thus it correctly predicts the presence of a low energy scale 
(proportional to the Heisenberg superexchange $J$) that regulates the N\'eel temperature 
at large $U/t$, as opposed to the scale $U/t$ for $T_N$ favored by the ``naive'' HF method.

The comparison with DQMC also provides additional evidence that the new technique 
is not only qualitatively correct, but it provides reasonably quantitative values 
for $T_N$. The DQMC data shown are for up to 10$^3$ clusters\cite{muramatsu-1} while 
our MCMF data are shown for 4$^3$ clusters via the red squares (results for larger 
lattices will be discussed below).  Both capture the $t^2/U$ scaling of $T_N$ at 
large $U/t$. At small $U/t$, $T_N$ tends to zero with decreasing $U/t$ consistent 
with the $T_N \sim \exp[-2\pi t/U]$ scaling derived from the weak 
coupling random phase approximation.\cite{wc2,wc1} The crosses are obtained 
from a finite-size scaling analysis of the results 
generated by the MCMF method, and represent $T_N$ in the 
thermodynamic limit. For numerical ease, we calculate 
the majority of the three dimensional data for 4$^3$ systems, so to be consistent 
we show prominently the 4$^3$ $T_N$ in Fig.~\ref{rf-1}. Finally, note that 
while the MCMF results are close to those of DQMC, the $T_N$ values are  
consistently underestimated in the present approach. Yet, qualitatively the
MCMF results are correct.
(Also note that Ref.~\onlinecite{muramatsu-1} contains results of previous DQMC studies 
and the trend is that the predictions for $T_N$ are consistently decreasing 
with time as the results are more refined.) 
Nevertheless, for the purposes of 
testing the method (and in anticipation of the fact that the important application of 
MCMF will arise for multi-orbital systems where semiquantitative information 
will be sufficient due to the absence of DQMC), this degree of accuracy 
is quite acceptable.
 
Another important feature missing in the standard finite-temperature 
HF approach is the presence of local moments \textit{above} $T_N$. 
The area shaded in blue in Fig.~\ref{rf-1} shows the region with  
preformed local moments found with MCMF: the gray-blue boundary 
demarcates the crossover between regions with and 
without preformed local moments (it is just a crossover because
the transition is smooth). The blue dashed line with triangles indicates 
the HF $T_N$, which also corresponds to the local moments formation 
in that crude mean-field approach.  The crossover temperature increases 
monotonically with $U/t$ for $U\ge 6t$. At large $U$ it follows the HF $T_N$. 
Similar agreement has been reported in two dimensional DQMC results.\cite{scalettar-1} 
The determination of the crossover temperature and its systematics  for $U< 6t$ is discussed below. 
\begin{figure}[t]
\centering{
\includegraphics[width=5cm, height=10cm, clip=true]{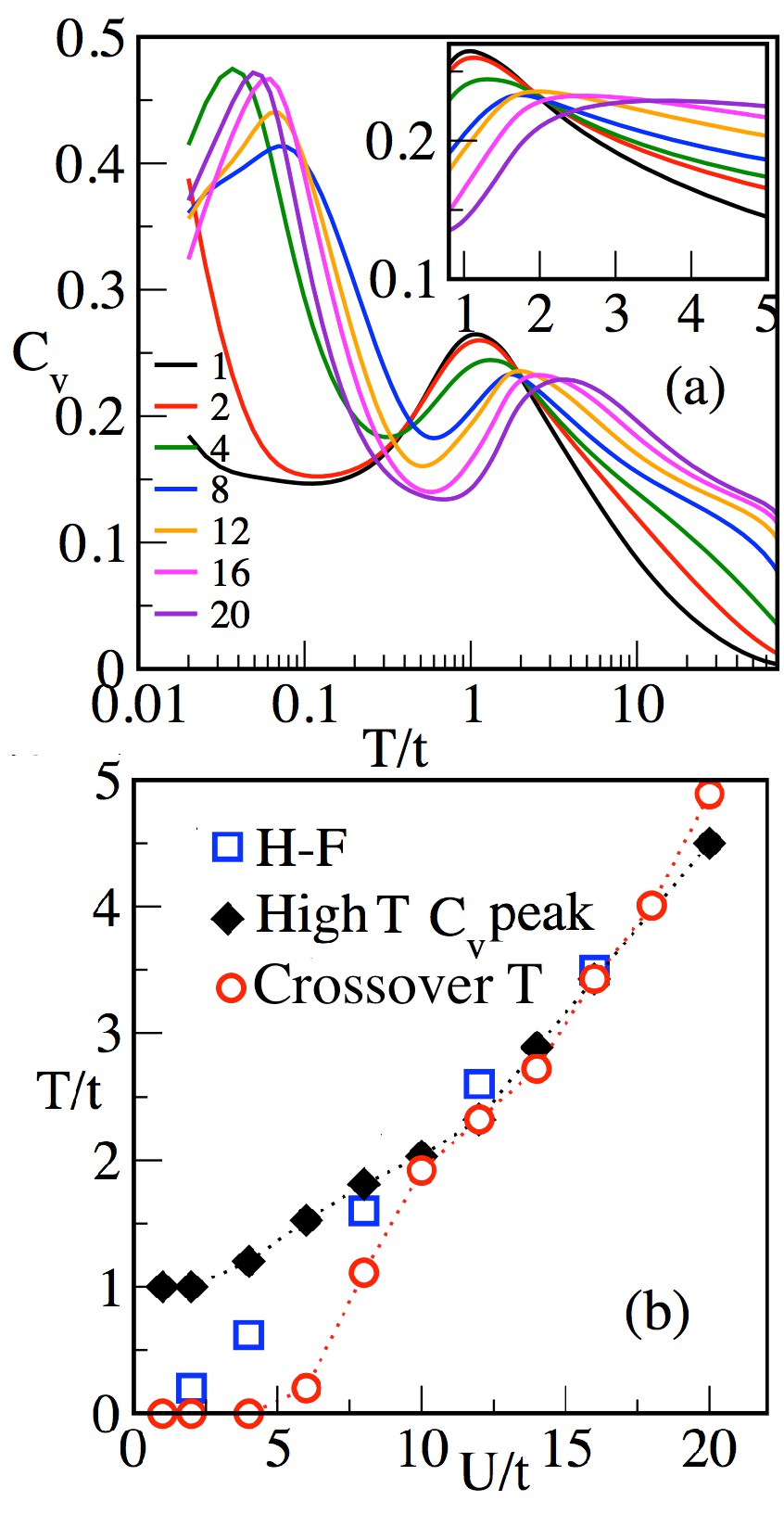}}
\caption{(color online)  (a) Specific heat vs. temperatures for different $U$ values. 
Two peak structures are observed: at large $U/t$ the high-temperature peak corresponds to the moment 
formation while the lower one to moment ordering. The inset shows the universal 
crossing of the different $C_v$ curves at $T/t\sim2.0$. 
(b) shows the position of the high-temperature peak
varying $U$. The Hartree-Fock results are shown with open squares. 
At large $U$, beyond $6t$,  these peak positions are close to the 
mean-field results. At low $U$, the high-temperature peak position 
saturates to $1t$, while the low-$T$ peak approaches zero. The non-merging 
of the two peaks in three dimensions is in agreement with DQMC data in 2D. 
The low-temperature peak correspond to $T_N$ in Fig.~\ref{rf-1}. 
The open circles in (b) show the crossover temperature from the 
no local moments regime to a region of preformed moments as obtained 
from the data on double occupation shown in Fig.~\ref{rf-2} (c). 
In (a) the data shown is a smoothed version of the actual data to reduce statistical fluctuations. }
\vspace{-0.0cm}
\label{rf-3}
\end{figure}
\textit{(i) Local moments and magnetic order:} 
Typical structure factors $S({\bf q})$ for ${\bold q}=(\pi,\pi,\pi )$ 
are shown in Fig.~\ref{rf-2} (a).  Here, we observe the non-monotonicity 
of $T_N$ with increasing $U$. This provide the finite size data for $T_N$ 
shown in Fig.~\ref{rf-1}. 
The moment formation vs. temperature is shown in panel (b). The system-averaged 
local moment is defined as $M=\langle (n_{\uparrow}-n_{\downarrow})^2\rangle = 
\langle n\rangle-2\langle n_{\uparrow}n_{\downarrow}\rangle$ with 
$\langle n\rangle=\langle n_{\uparrow}+n_{\downarrow}\rangle$. We 
note that for our rotation invariant case, 
$M=4\langle S_z ^2\rangle=4\langle ({\bold S}\cdot\hat{\Omega})^2\rangle$, 
where $\hat{\Omega}$ is an arbitrary unit vector.

For the half-filled $\langle n \rangle $=1 
uncorrelated case $U=0$,  $\langle n_{\uparrow}n_{\downarrow}\rangle =\langle
n_{\uparrow}\rangle\langle n_{\downarrow}\rangle$=1/4. Thus for $U=0$, or
alternatively $T/t \gg U/t$, $M=\frac{1}{2}$. 
This is seen in Fig.~\ref{2}(b) at $T/t\sim100$ for all values of $U/t$.  
From panel (c) we also observe that the average double occupation at high temperature 
for all $U/t$ values shown tends towards 0.25, the uncorrelated value of double occupancy.
On the other hand, for large $U/t$ and very low $T/t$,  the double 
occupation,  $\langle n_{\uparrow}n_{\downarrow}\rangle$,  is much 
suppressed and $M \sim 1$, i.e. the $U=\infty$ result.  
For any finite $U/t$ there is some finite double occupation and $M$ is always 
smaller that unity. Furthermore, since smaller $U/t$'s have larger double 
occupation, as shown in Fig.~\ref{2}(c), $M(T\sim0)$ monotonically decreases 
with reducing $U/t$, as shown in Fig.~\ref{2}(b). 

We also notice that $M(T)$ has some features at intermediate temperatures 
that evolve with $U/t$. In the intermediate temperature range, specifically 
between $T/t=1$ and $0.01$, we find two kinds of behavior. At small $U/t$,
 up to $U/t=4$, $M$ has a minima at $T/t\sim 0.2$ before reaching its absolute 
maximum at $T=0$. For larger $U$, the $M$ maxima lies at finite $T/t\sim 0.1$ 
and $\sim 0.5$, for $U/t=8$ and $16$, respectively. It is clear that for 
large $U/t$ the system can be approximated by a spin-1/2 Heisenberg antiferromagnet. 
Excitations at small but finite temperature that perturb the AFM order also suppress 
the virtual exchange due to the Pauli exclusion principle. This increases the degree 
of localization and promotes larger on-site moment size thereby pushing the maxima 
of  $M$ to finite temperature. The features seen at low temperature for small $U$ 
are correlated to the thermal evolution of the fields updated with MC, 
as discussed later. Note that similar observations were reported before 
in two dimensional DQMC studies,\cite{scalettar-1} increasing the evidence 
that MCMF captures the essence of the problem.

\textit{(ii) Specific heat:}  
The temperature evolution of the local moment in Fig. \ref{rf-2}(b) shows a continuous 
increase with decreasing temperature  
up to $T/t=1$. But this does not provide clear information on the 
crossover location between regimes with and without local moments. To address 
this issue, and to further test the MCMF method, we calculated the specific heat $C_v$ 
vs. temperature for different values of $U$. Here, it is expected 
that $C_v$ vs. temperature should have a two peak structure, the peak at high temperature  
corresponding to moment formation and the peak at low temperature corresponds to 
moment ordering at large $U/t$. 

In Fig. \ref{rf-3}(a) the specific heat vs. temperature is shown for a 4$^3$ system, 
where we find the expected two peak structure. The locus of the low-temperature peak 
corresponds well with the $T_N$ shown in Fig.~\ref{rf-1}. The high-temperature 
peak positions vs. $U/t$ are in  Fig. \ref{rf-3}(b). Here we also show the HF data 
with open squares.  Clearly, beyond $U/t=10$ the MCMF result coincides with the HF result.  
At lower values of $U$ (below 4$t$), the high-temperature peak appears to saturate to $T/t=1$. 
On the other hand, the low-temperature peak is suppressed to zero with low $T$. 
We note that we were unable to reliably carry out the numerical derivative 
below $T/t=0.02$, but the trend of the low-temperature peak shifting towards 
zero is apparent here and is also in  Fig. \ref{rf-1} (solid, red squares). 
\begin{figure}[t]
\centering{
\includegraphics[width=8cm, height=8cm, clip=true]{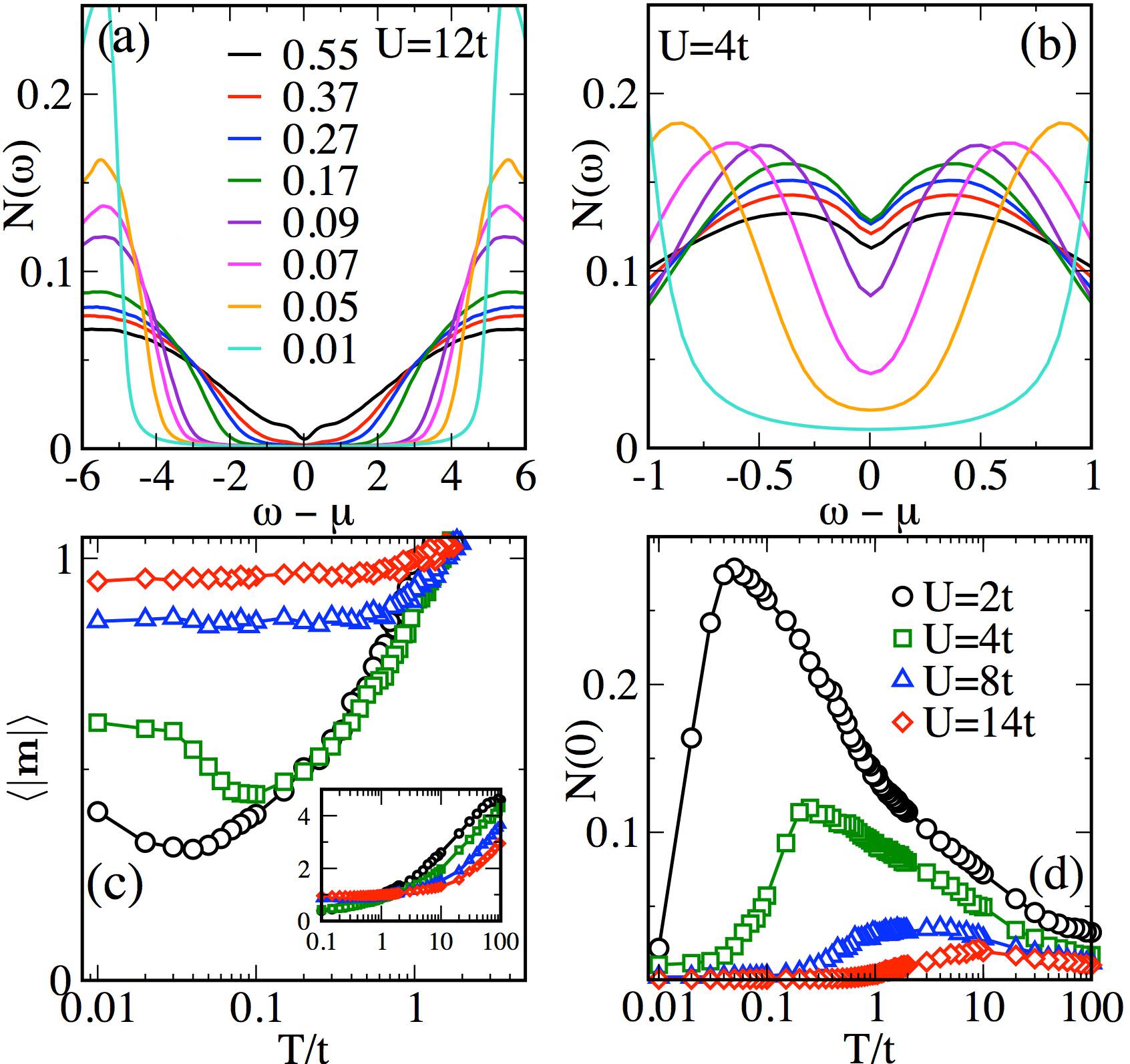}}
\caption{(color online) 
Density of states $N(\omega)$ for (a) $U/t=12$ and (b) $U/t=4$ at the temperatures 
indicated in panel (a). At large $U/t=12$, the Hubbard gap is gradually filled up 
due to thermal fluctuations. The weight at $\omega-\mu=0$ monotonically increases 
with increasing temperature. A DOS pseudogap is seen above $T_N\sim 0.2t$. 
At a smaller  coupling $U/t(=4)$, while the gap is filled similarly 
to the large $U$ case with the increase of temperature, above $T_N$ we observe 
a non-monotonicity in the dependence of the zero-energy weight with temperature.  
(c) Magnitude of the auxiliary classical fields averaged over the 
lattice ($\langle|\textbf{m}|\rangle$) vs. temperature for the 
$U$ values in (d). At large temperature the thermal fluctuations 
cause ($\langle|\textbf{m}|\rangle$) to grow linearly with 
temperature for all the $U$'s shown (inset).  The reason for this 
temperature dependence of $\langle|\textbf{m}|\rangle$ 
and its correlation with $N(\omega=0)$ is discussed in the text. 
(d) Shows the $N(\omega=0)$ feature remarked in panel (b) for 
different $U$ values. This non monotonicity was 
reported before in a DQMC study, see Ref.~\onlinecite{scalettar-1}.}
\vspace{-0.0cm}
\label{rf-4}
\end{figure}
Thus, in the present study we report that the high- and low-temperature peaks 
do not merge with reducing temperature at small $U/t$ in three dimensions. 
(This is also the case in two dimensions, which is discussed later). Previous studies 
have not agreed on this issue: dynamical mean 
field theory (DMFT)\cite{dmft-1,dmft-2,dmft-3} and Lanczos 
on one-dimensional chains\cite{ed} find the two peaks merging 
together with reducing $U$ while a DQMC study in two dimensions\cite{scalettar-1}, 
agrees with our conclusions.  Here we have extended the results to three dimensions.
 
 Another feature arising from the independence of the high-temperature 
entropy\cite{moreo,scalettar-1} is a universal crossing in $C_v$. In two-dimensional DQMC, 
this occurs at $T/t \sim 1.6$, with a spread in temperature of $\sim0.2t$. This has been 
observed in DMFT\cite{dmft-1,dmft-2,dmft-3} as well. We find a similar crossing 
both in three and two dimensions. In three dimensions the crossing is at 
approximately $T/t=2.0$ and has a small spread for low $U$ values, while 
at larger $U$ there seems to be a systematic increase 
to higher temperature with increasing $U$. This last conclusion was reported 
earlier as well.\cite{moreo} 
For our main purpose of testing the MCMF method, in two dimensions 
once again our results agree well with DQMC data, as discussed later.

\textit{(iii) Crossover temperatures:} 
At large $U$, the high-temperature peak of  $C_v$ corresponds to the moment 
formation.\cite{scalettar-1} Thus, this peak is an indicator 
of the local moment formation temperature. Below $U/t=10$, however, this high-temperature 
peak deviates from the linear behavior seen in Fig.~\ref{rf-3}(b) and eventually 
saturates to $T/t=1$. The approach of the peak location to $T \sim t$ 
at small $U$ indicates that considerable contribution to this peak comes 
from electron delocalization. For this reason, for $U/t<10$, the high-temperature peak 
cannot be used as a reliable indicator of local moments.  Thus, we use the double occupation, 
as plotted in Fig.~\ref{rf-2}(c), as an alternative indicator. To do so we need 
to choose a cutoff because the local moment formation is not abrupt but occurs with
continuity.  This cutoff is shown is Fig.~\ref{rf-2}(c) with the horizontal dashed line. 
For a given $U$, the temperature where double occupation goes below the cutoff 
is taken to be the crossover temperature to a region with preformed local moments. 
In principle the choice of such a cutoff is arbitrary, however, the $C_v$ calculation 
serves as a guide.
To address this issue, we chose a cutoff value such that the location of the 
high-temperature peak in temperature and the crossover temperature 
from the cutoff coincide at large $U$ ($=18t$). 
The crossover temperature for all other $U$ values 
are obtained from this fixed cutoff. They are plotted in  Fig.~\ref{rf-3}(b) 
with open circles. Clearly, there is  a good agreement with the high-temperature 
$C_v$ peak locations for large $U$. For $U/t<10$, we find a sharp deviation from 
linearity in the crossover temperature. As seen from Fig.~\ref{rf-2}(c), the 
crossover temperature for $U/t=6$ is very close to the corresponding $T_N$ 
in Fig.~\ref{rf-1}. For lower $U$ values, for this choice of cutoff, 
large double occupation considerably suppresses the local moment formation. 

\textit{(iv) Density of states:} 
In the half-filled Hubbard model, the charge gap is directly related 
to the existence of the local moments \textit{regardless} of magnetic order. 
This charge gap manifests as a gap (zero spectral weight in a finite energy range) 
in the DOS at $T=0$. With increasing temperature, this hard gap softens 
and is replaced by a pseudogap, with the spectral weight in the gap gradually 
increasing with increasing temperature. At large $U/t$ this monotonic behavior 
is seen in Fig.~\ref{rf-4}(a) from our MCMF results.  The DOS is displayed 
up to $T=0.55t$, but the monotonicity persists to higher temperatures. 
In contrast, 
at $U/t=4$, shown in Fig. \ref{rf-4}(b), the pseudogap spectral weight 
has a non-monotonic behavior: 
for $T>0.17t$ the spectral weight 
at $\omega-\mu=0$ decreases with increasing temperature, while 
for $T < 0.17t$ the spectral weight decreases with decreasing temperature.
\begin{figure}[t]
\centering{
\includegraphics[width=8cm, height=9cm, clip=true]{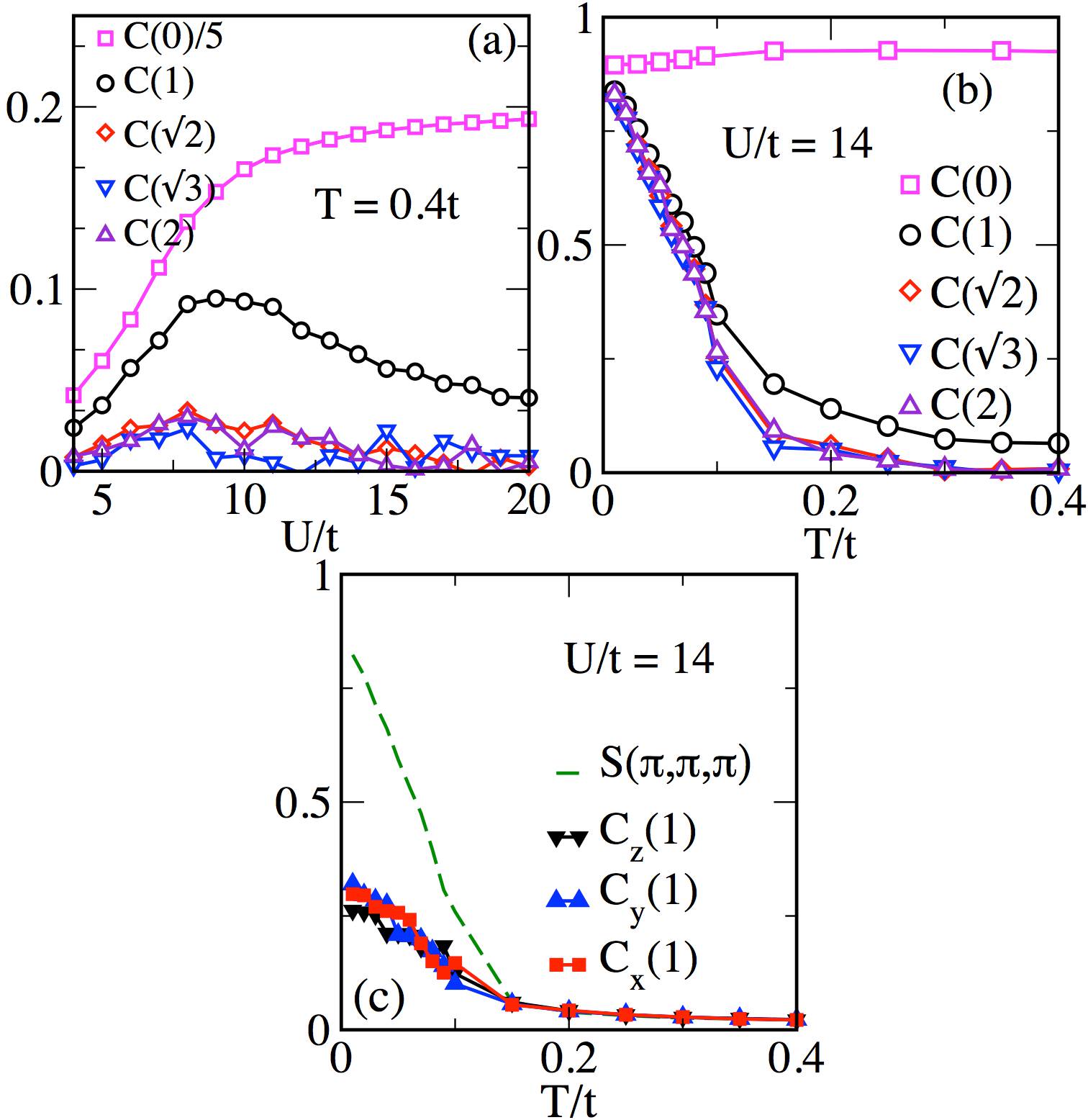}}
\caption{(color online)  (a) Real space spin-spin correlations $C(|{\bold r}|)$, for $|{\bold r}|=0,1,\sqrt{2},\sqrt{3},2$, at $T/t=0.4$, i.e. a temperature above $T_N$. See the text for the definition of $C(|{\bold r}|)$. The $|{\bold r}|=0$ curve corresponds to the square of the local moment and shows that the size of the preformed magnetic moment increases with $U/t$ and saturates 
beyond $U/t\sim 8$, i.e. where $T_N$ is maximized. The rest of the curves 
show the real space AFM correlations, ${\bold q}=(\pi,\pi,\pi$), among the moments.  
Again the correlations are the largest for $U/t\sim 8$. On the large $U$ side the 
decrease in the correlation results from thermal fluctuations competing with the 
AFM spin order stiffness which scales as $t^2/U$. (b) Shows the dependence 
of $C(|{\bold r}|)$ on temperature for large $U/t(=14)$. The magnitude of the moment is almost independent of the temperature, while there is a clear short-range AFM correlation between the moments at all temperatures shown. Longer range correlations for $|{\bold r}|>1$ are suppressed rapidly above $T_N\sim0.12 t$. (c) The real-space correlations  $C(|{\bold r}|=1)$  using only the $x$, $y$, or $z$ 
components of the spin. The data confirms explicitly the rotational 
invariance expected to exist in $H_{eff}$. The AFM structure factor is 
also displayed for comparison. Results are similar for $|{\bold r}|>1$ as well.}
\vspace{-0.0cm}
\label{rf-5}
\end{figure}
Since the spectral weight at $N(\omega=0)$ results from the scattering 
of the electrons from the classical fields, in Fig.~\ref{rf-4}(c) 
we show the evolution of the corresponding system-averaged auxiliary field 
values, $ \langle |{\bold m}| \rangle$. For $U/t=2$ and 4,  
$\langle|\textbf{m}|\rangle$ has minima at $T/t=0.05$ and $0.1$, respectively. 
For $U/t=8$ and 14, $\langle|\textbf{m}|\rangle\sim1$ for $T/t\lesssim $ 0.5. 

The behavior shown in Fig. \ref{rf-4} 
can be explained as follows. At high enough temperatures 
with negligible local moments, the value of  $\langle|\textbf{m}|\rangle$ 
is governed by thermal fluctuations. At these temperatures the auxiliary fields 
behave as harmonic oscillators with a mean amplitude proportional to $\sqrt{T/U}$. 
Thus, $\langle|\textbf{m}|\rangle$ grows with increasing temperature.  On the other 
hand, the $T=0$ equilibrium value of $\langle|\textbf{m}|\rangle$  is directly 
proportional to $U$, as seen in Fig.~\ref{rf-4}(c). For smaller values of $U$, thermal 
fluctuations dominate to low enough temperatures, causing $\langle|\textbf{m}|\rangle$ 
to reduce to values smaller than their $T=0$ value. On further reduction in temperature, 
these thermal fluctuations are suppressed and $\langle|\textbf{m}|\rangle$  starts to increase 
towards its mean value at $T=0$. The minima in $\langle|\textbf{m}|\rangle$ vs. 
temperature corresponds to the location of the maxima in $N(0)$ in Fig.~\ref{rf-4}(d) 
for $U/t=2$ and 4. This indicates that the $N(0)$ suppression 
at high temperature results  from the scattering of electrons from 
thermally fluctuating large  $\{ {\bold m_i} \}$ fields, while at small 
temperatures, the reduction in $N(0)$ results from the depletion of 
spectral weight due to the opening of the
Mott gap.  The peak in the $N(0)$ occurs between the two regimes.

With increasing $U$, the dominance of thermal fluctuations 
in governing $\langle|\textbf{m}|\rangle$ is pushed to progressively 
higher temperatures as is also seen from the peaks of $N(0)$ for $U/t=8$ 
and 16 in Fig. \ref{rf-4}(d). At these temperatures $\langle|\textbf{m}|\rangle$ is higher 
than their $T=0$ values, thus no minima is found for these cases 
in Fig. \ref{rf-4}(c).

We stress that the high temperature increase in the auxiliary field 
magnitude does not imply an increasing magnetic moment. As seen in Fig.~\ref{rf-2}(b) 
the magnetic moment $M$ saturates at its uncorrelated value of 1/2 at high temperature. 
The non-trivial effect of the fluctuations in the auxiliary fields is in the DOS, 
in the low-temperature feature in $M$  at small $U$, and possibly in the conductivity.

Another feature observed in the inset of Fig.~\ref{rf-4}(c) is that the magnitude 
of the auxiliary fields vs. temperature for different $U$'s cross 
between $T/t=1$ and $T/t=2$. Since at  large $T$, $\langle|\textbf{m}|\rangle$  grows 
as $\sqrt{T/U}$, the  auxiliary fields magnitude for smaller $U$ grows more rapidly 
than those for larger $U$. At small temperatures, however, the 
$\langle|\textbf{m}|\rangle$  values are directly proportional to $U$ 
as discussed before, naturally 
explaining the observed crossing. Note that this crossing coincides 
with the universal crossing of the specific heat in Fig.~\ref{rf-3}(a).

\textit{(v) Real space spin correlation: } 
Figure~\ref{rf-5}(a) shows the spin-spin correlation $C(|{\bold r}|)$ at
$T/t = 0.4 > T_N/t$ for different values of $|{\bold r}|$. The special case
$|{\bold r}|=0$ corresponds to $M$ and with increasing $U/t$, $C(|{\bold r=0}|)$
saturates. The most prominent real space AFM correlation at this temperature is
for $C(|{\bold r=1}|)$. 
While it is almost zero for $U/t \le 4$, it
increases as a function of $U/t$, reaches a maximum at $U\sim 8t$, 
and then reduces with further increases in $U/t$.  
The large $U/t$ suppression is due to the $t^2/U$ suppression of the spin ordering 
stiffness.  While a similar trend is seen for larger $|{\bold r}|$, the magnitude 
of the correlation is greatly suppressed. In Fig.~\ref{rf-5}(b) we show the evolution 
of $C(|{\bold r}|)$ with temperature at a typical large value of $U/t$.  
While the magnitude of the moment, given by $C(|{\bold r=0}|)$,  
increases slightly with temperature, the short-range 
correlations are suppressed rapidly beyond $T_N$. 
The increase in $C(|{\bold r=0}|)$ or 
the size of the local moment were discussed earlier. 
Only $C(|{\bold r=1}|)$  is robust above $T_N$. Finally 
in Fig.~\ref{rf-5}(c)  we  also show individually the $x$, $y$, and $z$ 
components of $C(|{\bold r}|)$ for $|{\bold r}|=1$ for $U=14t$. 
This confirms explicitly the rotational invariance of the calculation. 
\begin{figure}[t]
	\centering{
		\includegraphics[width=8.6cm, height=7.6cm, clip=true]{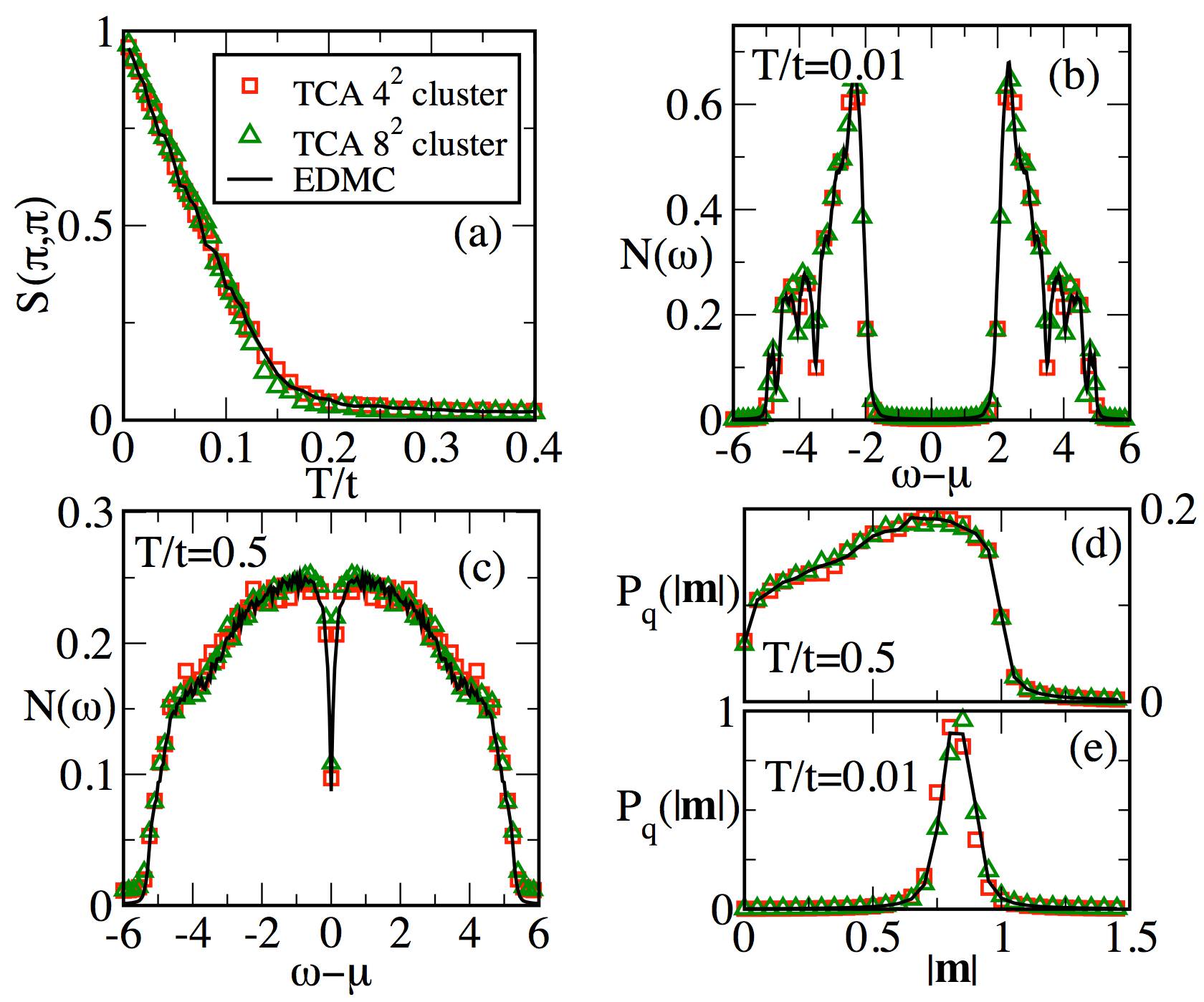}}
	\caption{(color online) Comparison of results for an 8$^2$ system at $U/t=6$, 
		obtained using ED+MC (solid line) and TCA with two different sizes of traveling cluster sizes, $4^2$ (squares) and $8^2$ (triangles). (a) shows $S(\pi,\pi)$, (b) and (c) 
		show the DOS, while (d) and (e) display 
		the $P_q(|{m}|)$,  for the three cases. The DOS and $P_q(|{m}|)$ are 
		shown at low $T(=0.01t)$ and high $T(=0.5t)$ temperatures, as indicated.}
	\vspace{-0.0cm} 
	\label{rf-6}
\end{figure}
\begin{figure}
\centering{
\includegraphics[width=8.6cm, height=4cm, clip=true]{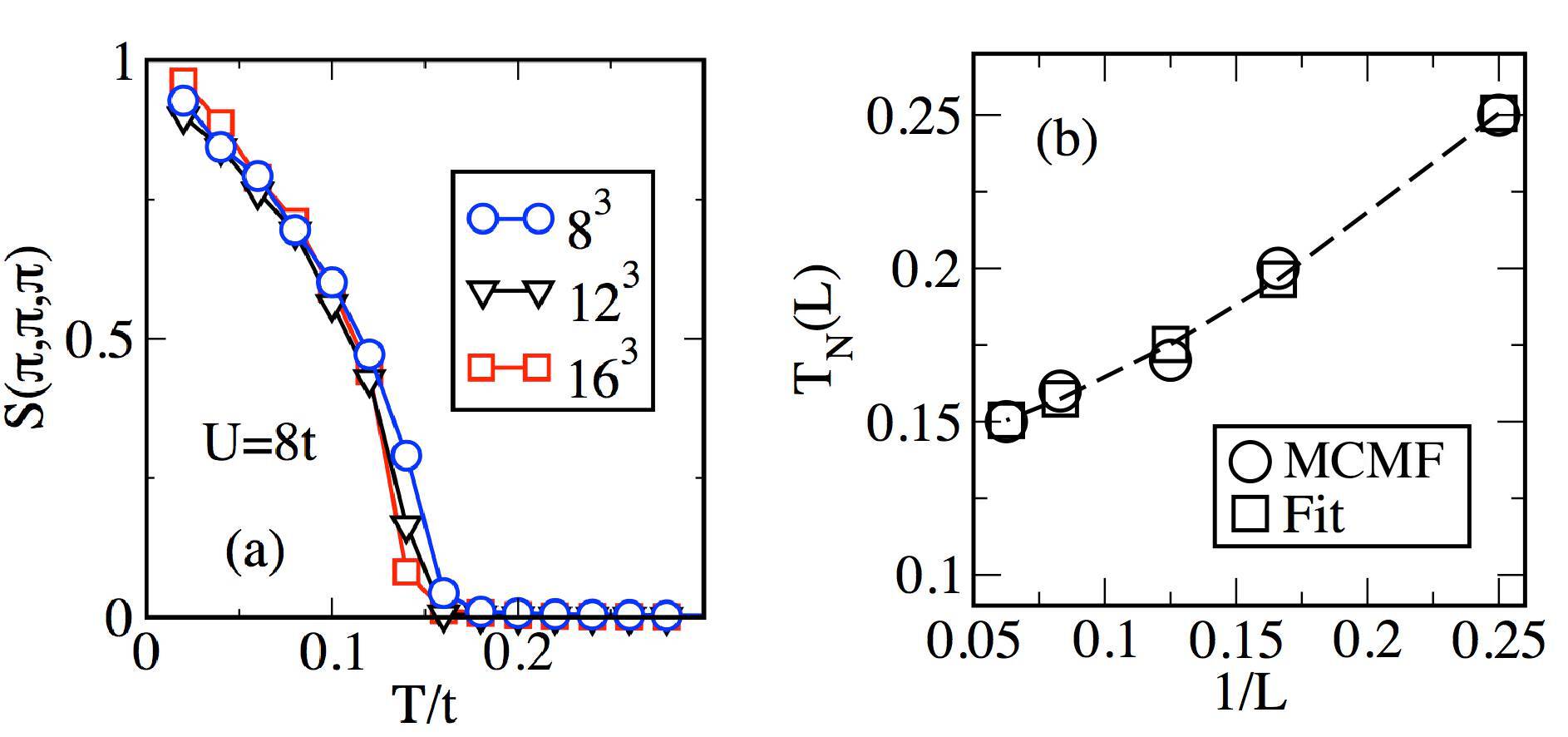}}
\caption{(color online) Finite size scaling analysis for the 3D case.  
(a) shows $S(\pi,\pi,\pi)$ for $U/t=8$ for three system sizes obtained using TCA. To find $T_N$ in the thermodynamic limit we fit $T_N(L)$, the N\'eel temperature for a cluster of size $L^3$, against 1/$L$. Fitting to a scaling form (see text) provides $T_N^{Thermo}$, the N\'eel temperature in the thermodynamic limit. As a typical example, in (b) we show the MCMF  $T_N(L)$ data and the fit using the scaling form for $U=8t$ . The dashed line is a guide to the eye. See text for discussion.}
\vspace{-0.0cm} 
\label{rf-7}
\end{figure}

Summarizing this subsection, we have established that the non-monotonic 
dependence of $T_N$ on $U$, the physics of preformed local moments, and 
the pseudogap features in the DOS can all be captured within the MCMF method.  

\subsection {Accessing larger system sizes}
ED+MC is numerically expensive since an exact diagonalization must be performed 
at every step in the process. The numerical cost of a sequential sweep 
scales as $O(N^3) \times N$, with $N$ the total number of lattice sites. 
To overcome this $O(N^4)$ scaling we employ a recently developed variation 
of real space ED+MC that scales \textit{linearly} with  
the system size.\cite{method-5} This technique, called 
``Traveling Cluster Approximation'' (TCA), defines a region (the traveling cluster) 
around the site where a MC update is attempted. 
A change is proposed and the 
update is accepted or rejected based on the energy change computed within 
the traveling cluster, thus bypassing the 
costly diagonalization of the full system.  
Only when observables are calculated, after equilibrium has been 
reached, is a full system diagonalization 
performed. This adds only a few hundred full system 
diagonalizations to the computational cost. For TCA, the computation 
cost of ED for a system with $N$ sites is $O(N_c^3 )$, where $N_c$ 
is the traveling cluster size. The cost of a full sweep of the lattice is $Nc^3\times N$ 
or linear in $N$ as opposed to $N^4$. This allows us to solve much 
larger systems. We now discuss our benchmarks for the TCA and the results 
on large two and three dimensional lattices.

\textit{1. Benchmarking:} Let us begin by comparing the results 
from TCA with ED+MC. 
In Fig.~\ref{rf-6} we compare various observables on two-dimensional 
8$^2$ clusters with periodic boundary conditions. Results are shown for 
two different traveling cluster sizes, namely $N_c = 4^2$ (squares) 
and $N_c = 8^2$ (triangles), while results for the full ED + MC are 
given as the solid lines. 
Fig. \ref{rf-6}(a) shows $S({\bf q})$ for ${\bold q}=(\pi,\pi)$, 
Figs. \ref{rf-6}(b)-(c) show the DOS, 
and Figs. \ref{rf-6}(d)-(e) show $P_q(|{\bold m}|)$
at low ($T=0.01t$) and high ($T=0.5t$) temperatures. 
All of the results are for $U=6t$. 
The $T_N$ obtained is about $0.15t$ from  
both methods. The Mott gap is found to be about $4t$ at low temperature 
[see \ref{rf-6}(b)]. The pseudo gap feature at $T/t=0.5$ in 
Fig. \ref{rf-6}(c) is also captured 
very accurately within TCA. Finally, all sites at low temperature 
show ${\bold m_i} \sim 1$ which evolves into a broad distribution at high temperatures. 
We find a satisfactory agreement between ED+MC and TCA data 
for all the observables and at all temperatures. Furthermore, 
these results show that employing a 4$^2$ traveling 
cluster is adequate as the results are virtually indistinguishable from those obtained 
using a 8$^2$ traveling cluster. In the following we will employ 4$^2$ 
and 4$^3$ traveling clusters in two and three dimensions, 
respectively. 

\textit{2. Finite size scaling:} With TCA-based MCMF 
now we can study up to 16$^3$ lattices. 
As a result the $S(\pi,\pi,\pi)$ data is available 
for $N= 4^3$ to $16^3$ system sizes. 
Moreover, the magnetic structure factor obtained with the TCA agrees with 
the ED+MC data at all temperatures. This indicates that finite size effects 
associated with the TCA do not affect the finite temperature evolution of the 
magnetic state. 
Hence we can employ a finite size scaling analysis to the TCA 
data in order to obtain 
the N\'eel temperature in the thermodynamic limit.

On a finite system, estimates of  T$_N$ can be obtained either from 
an inspection of the $S(\pi,\pi,\pi)$  data or from the maxima of 
thermodynamic quantities such as the specific heat or the magnetic 
susceptibility. Then, assuming that the correlation length $\xi(T_N(L)-T_N^{Thermo})=aL$ 
on a $L^3$ system, and given that $\xi(x)\propto|x|^{-\nu}$, 
one arrives at the scaling form, 
$T_N(L)=T_N^{Thermo}+bL^{1/\nu}$. Here, $L$ denotes data from a $L^3$ cluster size.
We plot the finite cluster N\'eel temperatures against $1/L$ and use $T_N^{Thermo}$, $b$, and $\nu$ as fit parameters.  
A typical data fit is presented in Fig.~\ref{rf-7}(b). 
For reference, we provide the $S(\pi,\pi,\pi)$ data for different system sizes 
in Fig.~\ref{rf-7}(a). 
The crosses in Fig.~\ref{rf-1} are $T_N^{Thermo}$ obtained from this finite 
size scaling analysis. 

\begin{figure}[t]
	\centering{
		\includegraphics[width=8.6cm, height=4.6cm, clip=true]{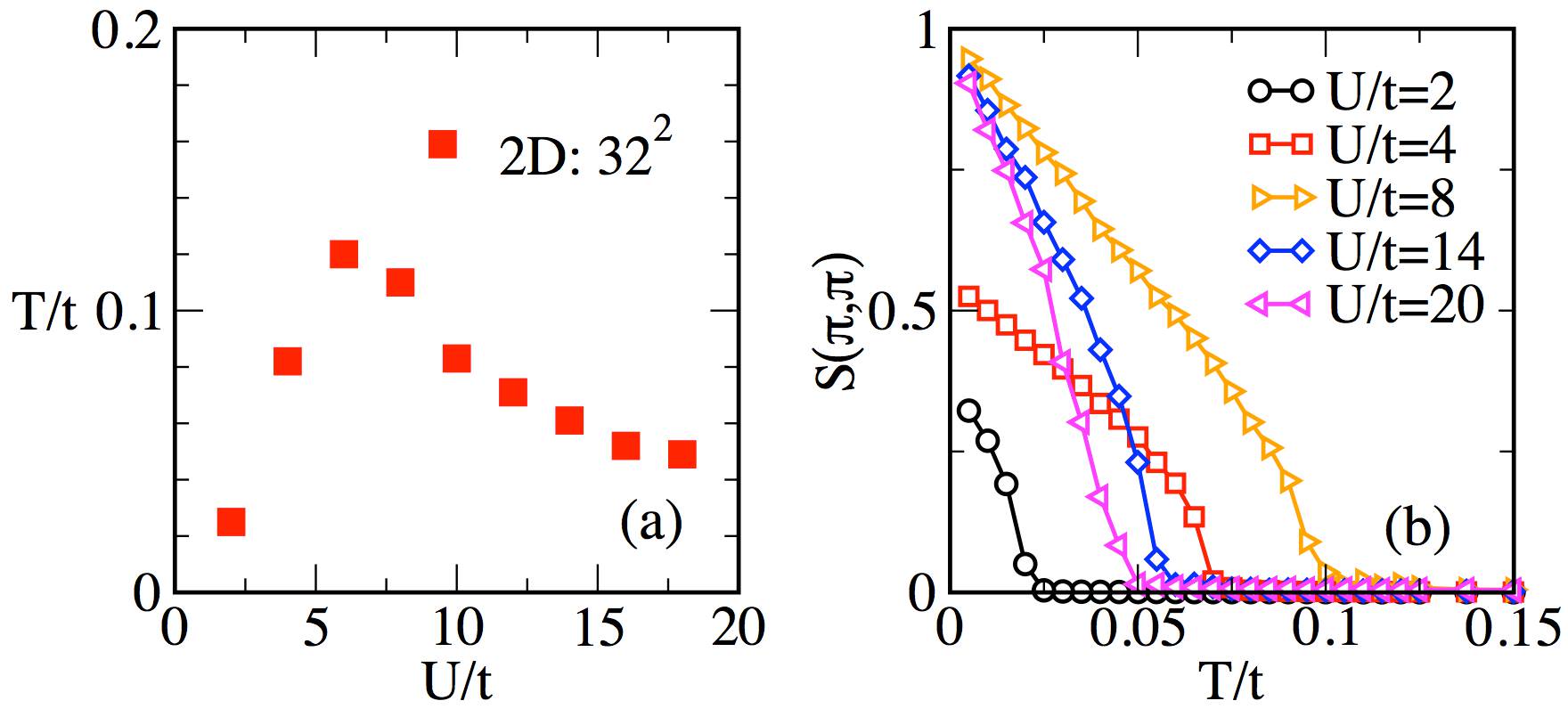}}
	\caption{(color online) (a) $T_N$ vs $U/t$ and (b) 
		representative $S(\pi,\pi)$'s, both for a 32$^2$ system. 
		The results are obtained using a 4$^2$ traveling cluster.}
	\vspace{-0.0cm} 
	\label{rf-8}
\end{figure}

 \begin{figure}[t]
 	\centering{
 		\includegraphics[width=6.5cm, height=10.5cm, clip=true]{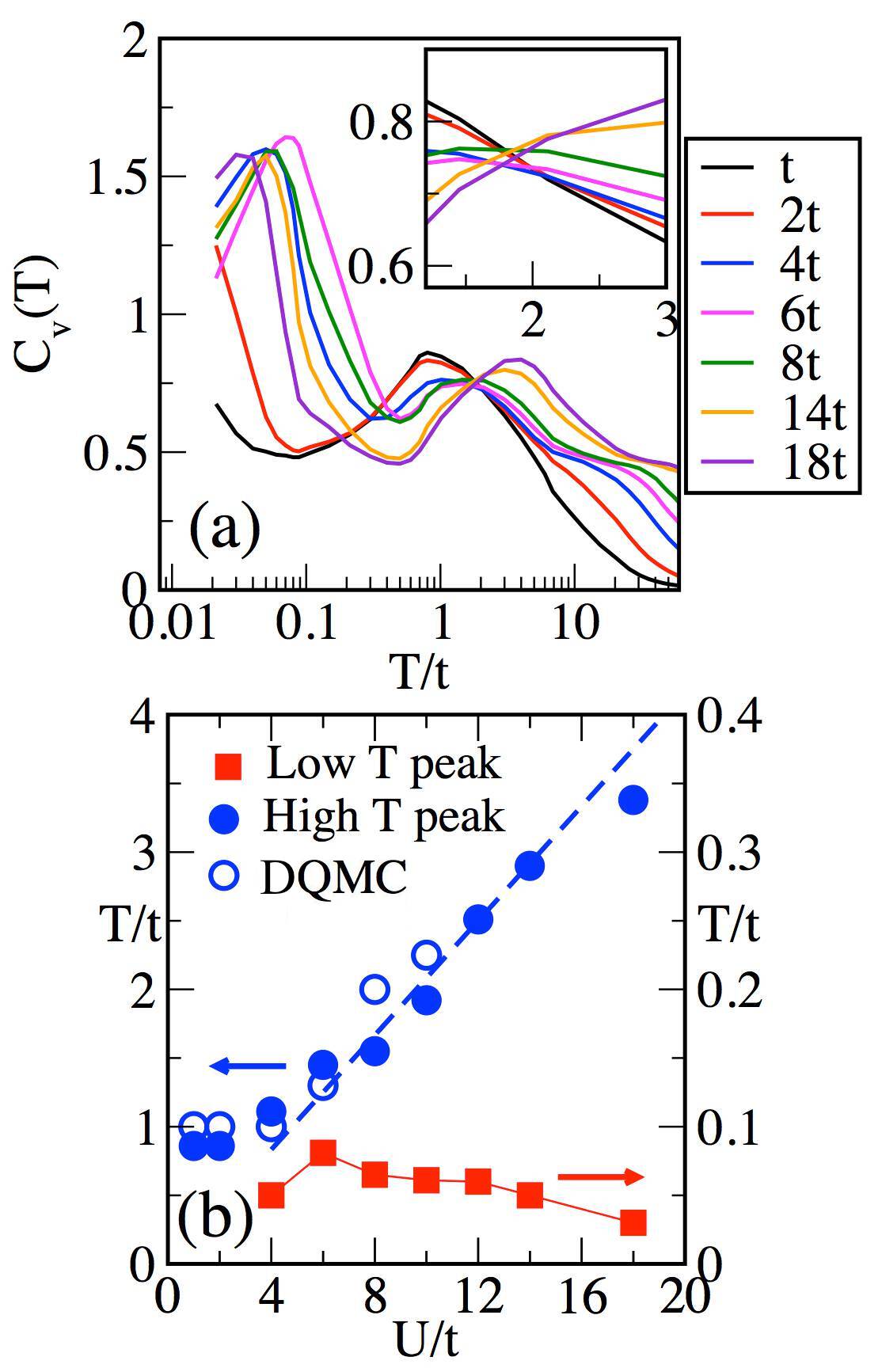}}
 	\caption{(color online)  The specific heat vs temperature data in two dimensions for various $U$ values. The twin peak structure and universal crossing are clearly seen. The corresponding loci of the high- and low-temperature peaks  are shown in (b). Here we also show DQMC 
data for the high-temperature peaks. 
For comparison  with DQMC, we show this data for a 6$^2$ system size. The dashed line is a guide to the eye.}
 	\vspace{-0.0cm} 
 	\label{rf-9}
 \end{figure}

\subsection {The two-dimensional lattice}
We now turn to results for large two-dimensional system sizes using a 4$^2$ traveling cluster. 
The results shown in Fig.~\ref{rf-8} are for a 32$^2$ system. The method can typically 
be pushed up to 40$^2$ sizes. Fig. \ref{rf-8}(a) shows the AF N\'eel temperature. 
Note that in principle the  
Mermin-Wagner theorem establishes that there is no true $T_N$ in two dimensions 
for an $O(3)$ magnet.  
This theorem is valid only for short-range spin-spin interactions, however. 
In our case, the integration of the fermions leads to effective spin-spin interactions at 
all distances, although the rate of the decay of the couplings with distance is unknown. 
To be cautious we should refer to this scale as $T_{corr}$ instead of $T_N$,
but below we will continue using the $T_N$ notation for this temperature scale since this
is the convention widely used in the literature.
Here, we observe that $T_N$ has the correct 
scaling of $t^2/U$ at large $U$. 
The corresponding spin structure factors are shown in Fig. \ref{rf-8}(b).

To make a comparison with two dimensional 
DQMC data,\cite{scalettar-1} $C_v(T)$ as well as the locus of the high- and low-temperature 
peaks of the specific heat are shown in Fig.~\ref{rf-9}. As in the three dimensional case, 
we observe a two-peak structure in the specific heat and also capture the universal crossing 
of the $C_v(T)$ for different $U$ values in (a).  The crossing occurs at $T/t=1.6$ and has a 
small spread in temperature values. These are in agreement with the DQMC data for the same system size. 
In \ref{rf-9}(b) we show the comparison between our data and the peak locations in DQMC. 
We find that the saturation of the high-temperature peak ($T_{high}$)  for $U/t<4$, 
persists in two dimensions. The small $U$ saturation of $T_{high}$ can be understood 
by studying the $U=0$ limit, where the specific heat peaks at $T\sim t(=1)$. For the 
behavior at large $U$ we can analyze the limiting case of $t=0$ (single site problem), 
where it is easy to see that $T_{high}$ grows linearly proportional to $U$. 
This linear growth of $T_{high}$ at large $U$ is seen in  Fig.~\ref{rf-9}(b), 
and is in good agreement with DQMC results.\cite{scalettar-1}

\begin{figure}[t]
 	\centering{
 		\includegraphics[width=9cm, height=4.5cm, clip=true]{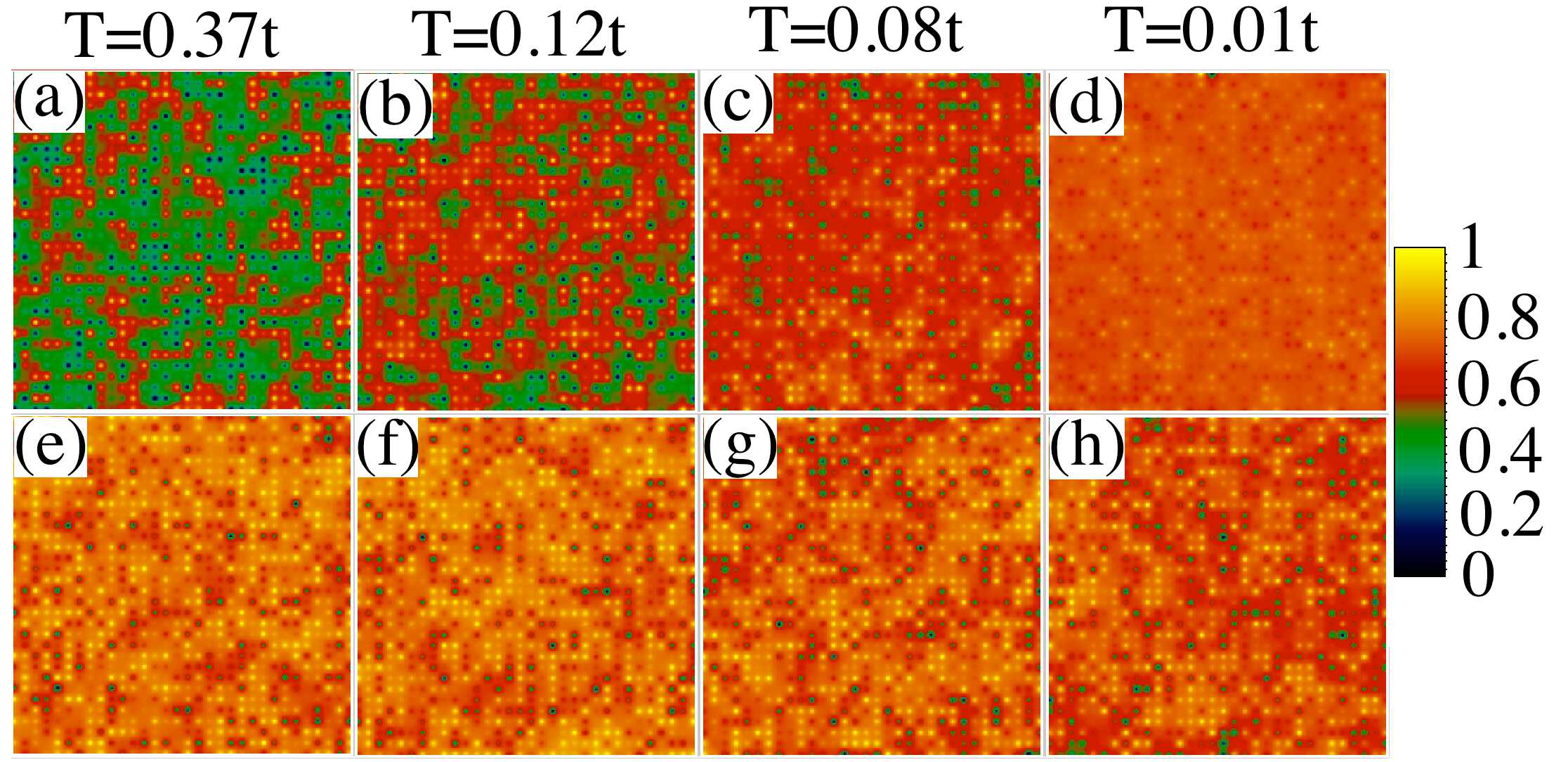}}
 	\caption{(color online) Spatial snapshots of $\{|{\bold m_i}|\}$ 
for $U/t=4$ (top) and $U/t=14$ (bottom). (a)-(d) and (e)-(h) show the 
snapshots for $T/t$=0.37, 0.12, 0.08, and 0.01 for the two cases, respectively.  
 		At $U/t=14$ (bottom), 
values as high as $|{\bold m}|\sim 1.0$ exist at all temperatures shown, 
much above $T_N (\sim 0.06t)$. $|{\bold m}|$, however, 
grows with reducing  temperature and shows thermal fluctuations at the higher temperatures 
for $U/t=4$  (top). In the figure, yellow implies $|{\bold m}|= 1$ and black $|{\bold m}|= 0$. 
The snapshots are for a $32^2$ system size.}
 	\vspace{-0.0cm} 
 	\label{rf-10}
 \end{figure}

The large system sizes accessible to MCMF 
allows a detailed analysis of the spatial evolution with temperature
of the $\lbrace {\bold m_i} \rbrace$ field configurations, 
as shown in Fig.~\ref{rf-10}. Here,  
the top and bottom panels contain the spatial maps of $|{\bold m_i}|$
for $U/t=4$ and $14$, respectively. The maps are shown for 
four different temperatures, decreasing from left to right. 

The temperature range here was chosen to
show that in the small $U$ case the $\lbrace {\bold m_i}\rbrace$ 
grows with decreasing temperature, similar to the case in 
three dimensions. The strong thermal fluctuations that 
make the MCMF approach accurate at high temperatures 
are clearly visible. At $U/t=4$, the magnitude of $|{\bold m}|$ 
has a broad distribution, with regions of small and large values 
[see Fig. \ref{rf-10}(a)]. At temperatures above but close 
to $T_N \sim 0.08t$, regions with $|{\bold m}|\sim 0.7$ start 
spanning the entire system, as exemplified in Figs. \ref{rf-10}(b) and 
\ref{rf-10}(c).   
Fig. \ref{rf-10}(d) shows the system below $T_N$.
These thermal fluctuations imply fluctuating 
spin moments in a MC snapshot, however, averaging over spin moments from many 
such MCMF configurations, at a fixed temperature, 
results in moments that are uniform in space. We stress 
that there is \textit{no} spatial phase separation implied in these snapshots.

\begin{figure}[t]
	\centering{
		\includegraphics[width=9cm, height=4.5cm, clip=true]{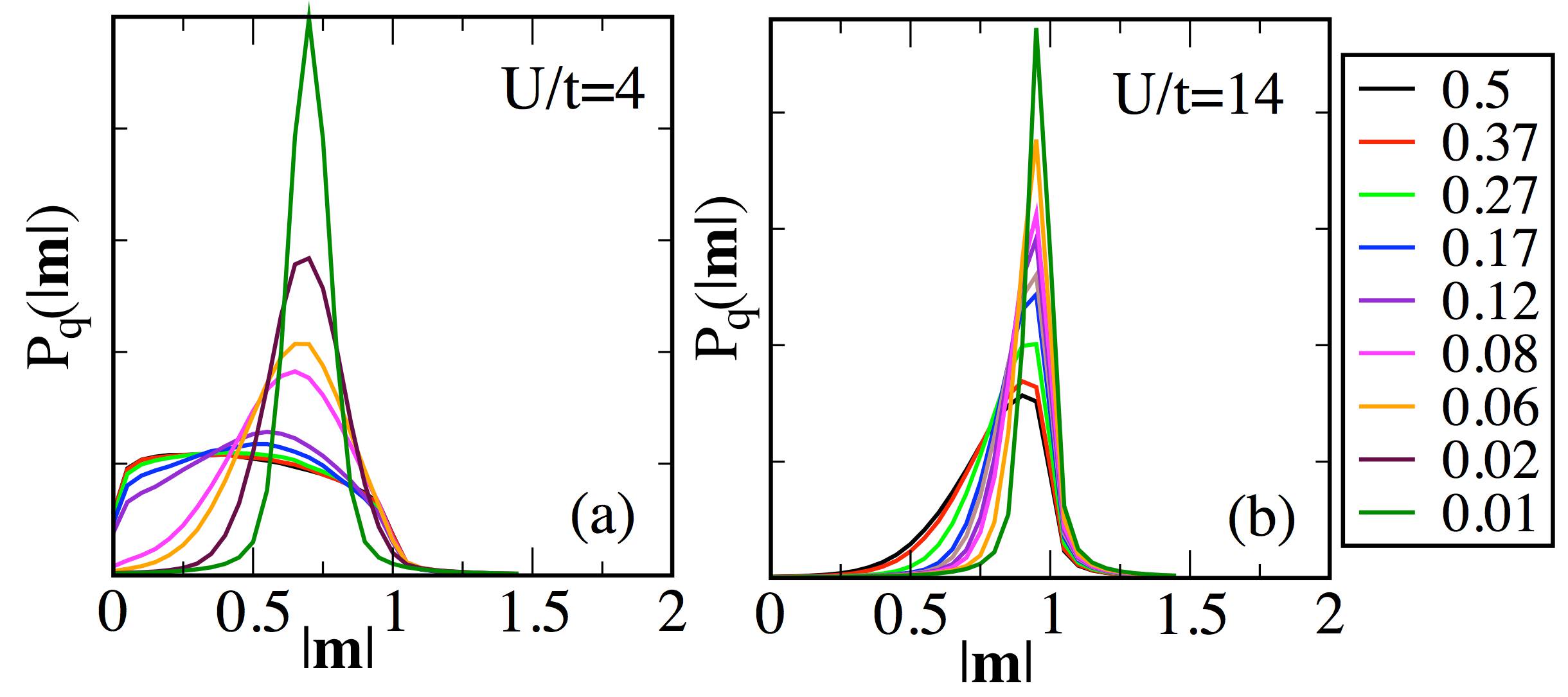}}
	\caption{(color online) $P_q(|{m}|)$ for $U/t=4$ and 14 at various temperatures indicated on the right. (a) At $U/t=4$ and high temperature, lattice sites acquire values of $|{\bold m}|$ between 0 and 1 in a uniform manner. This distribution starts peaking at $T/t\sim0.08$ which is close to $T_N$.  
		At lower $T$, the auxiliary field distribution peaks at about $|{\bold m}|=$0.7. 
		(b) At large $U/t$, the $|{\bold m}|$ values are well defined moments (about 1) 
		at all sites at the temperatures shown, much above $T_N$. There is only a small thermal broadening even at $T\sim 10T_N$.
	}
	\vspace{-0.0cm} 
	\label{rf-11}
\end{figure}

The corresponding distribution of the $\lbrace {\bold m_i}\rbrace$ 
configurations for Fig.~\ref{rf-10} is shown in Fig.~\ref{rf-11}. In Fig. \ref{rf-11}(a), we observe 
a gradual increase in the sharpness and peak height of $P_q(|{m}|)$ 
with reducing temperature. At large $U$, $P_q(|{m}|)$ shows little 
thermal  fluctuations in the temperature range shown. This corresponds 
to almost saturated $\langle {\bold |m|} \rangle$ as in three 
dimensions at temperatures below $T/t\sim1$. 
The uniformity in the bottom panel of Fig.~\ref{rf-10} 
translates into a sharp $P_q(|{m}|)$ in Fig.~\ref{rf-11} 
for a typical large values of $U/t$, 
as shown in Fig.~\ref{rf-11}(b) for $U/t = 14$.

\subsection{The half-filled Hubbard model with longer-range hopping} 

In this section we extend our analysis and apply the MCMF method 
to study the Hubbard model on a two-dimensional square lattice 
with nearest neighbor and next-nearest-neighbor hopping  
$t$ and $t^\prime$, respectively. 
The $t^\prime$ hopping processes have been widely 
considered important in the context of the cuprate superconductors, both directly in the 
Hubbard model,\cite{uttp-1} as well as in the $t-t^\prime-J$ model.\cite{uttp-2}  
In addition, understanding the role of $t^\prime$ is in general relevant to the study 
of frustrated systems. For this model DQMC studies, suffer a 
severe fermion sign problem due to the broken particle-hole symmetry 
introduced by $t^\prime$. Thus, the ground state 
properties remain inaccessible. DMFT studies have had more success 
but with limited or no spatial correlations.\cite{uttp-3} 
There are other approaches to access the ground 
state properties,\cite{gs-1,gs-2,gs-3} but they are difficult 
to generalize to finite temperature.  
MCMF can fill this void.

The MCMF approach used here reduces to unrestricted Hartree-Fock at $T=0$, 
but, as shown by comparison with DQMC results earlier, it rapidly improves 
its accuracy with increasing temperature. Moreover, MCMF does not have a 
sign problem.
Thus, it allows controlled calculations of both finite temperature 
and ground state  properties on very large two and three 
dimensional clusters, under a broad variety of circumstances. 
With this in mind, here we address the $U-t-t^\prime$ model using MCMF. 
We also use DQMC to solve the same problem 
for the lowest temperature allowed by the sign problem. 
\begin{figure}[t]
	\centering{
		\includegraphics[width=8.5cm, height=8.5cm, clip=true]{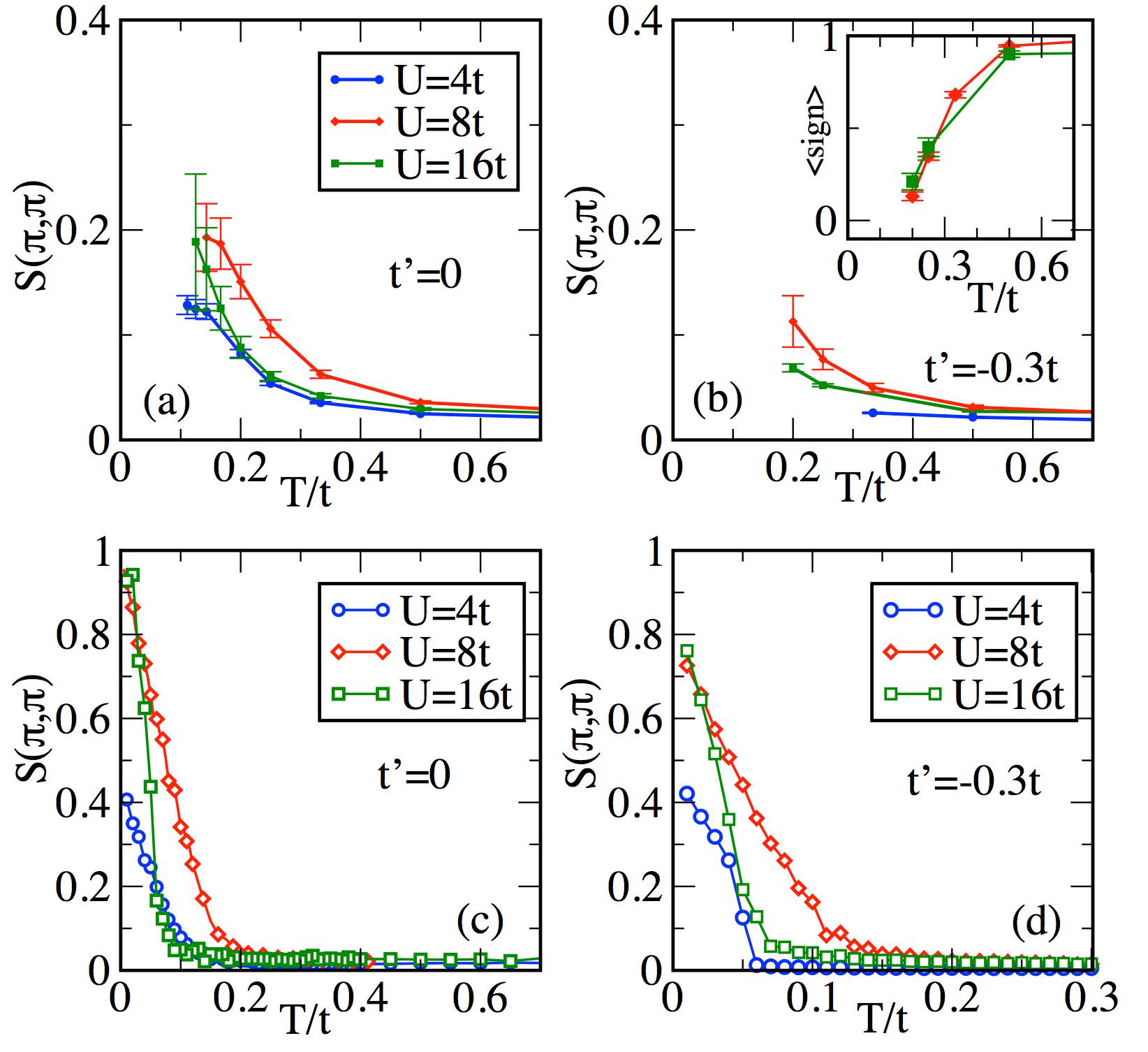}}
	\caption{(color online) DQMC results showing $S(\pi, \pi)$ for several $U$'s
		at (a) $t^{\prime}=0$ and (b) $t^{\prime}=-0.3t$. The inset in (b) contains 
		the average sign vs. temperature. For $t^\prime = -0.3t$, sign error increases rapidly 
		preventing the access to low temperatures. (c) and (d) show the $S(\pi, \pi)$ obtained from MCMF. The data shown here is for a 6$^2$ system. }
	\vspace{-0.0cm} 
	\label{rf-12}
\end{figure}
\textit{1. Comparison with DQMC :} In Figs.~\ref{rf-12}(a) and \ref{rf-12}(c), 
we show $S(\pi, \pi)$ calculated using DQMC and MCMF, respectively, 
for $t^\prime=0$. For this case, DQMC does not have sign problems and 
in principle we could obtain results for lower temperatures. However, 
given the $(O(L)^2)$ scaling in CPU time (where $L$ is the number of 
imaginary time slices\cite{WhitePRB1989}) and 
the existence of results in the literature, we stopped the DQMC calculation 
at $T/t\sim 0.1$.  
It is clear that even at these temperatures we do observe 
AFM correlations beginning to grow with reducing $T$. 
We also observe that magnetic correlations begin to grow at a higher temperature 
for $U/t=8$ compared to $U/t=$ 4 and 16. This is indicative 
of the non-monotonicity of $T_N$ with $U$, as extensively discussed earlier. 
By comparison, MCMF ordering happens at a lower temperature. 
An additional difference 
with DQMC is the high temperature tail seen in \ref{rf-12}(a), which is 
absent in \ref{rf-12}(c). 
As shown for three dimensions in Fig~\ref{rf-5}(b), however,  
short range spatial correlation, in particular C($|{\bold r}=1|$), survives 
up to high temperatures. Similar correlations survive in two dimensions as well, 
but the presence of quantum effects makes the AFM 
correlations survive to longer length scales in DQMC contributing to 
the high temperature tail. In contrast, since only C($|{\bold r}=1|$)  is significant 
in MCMF, the magnetic structure factor in \ref{rf-12}(c) has a suppressed tail. 
This comparison highlights the effect of the mean field approximation in MCMF at low $T$ 
on long-range correlations and may explain the reduced values of $T_N$ as compared with DQMC.

In Fig.~\ref{rf-12}(b) we present $S(\pi, \pi)$ for the physically relevant 
case $t^\prime/t=-0.3$.\cite{RMP94} In the inset, we show the 
average value of the fermion sign as a function of temperature. 
The loss of particle-hole 
symmetry causes the average sign to rapidly fall to zero. 
As a result it becomes impossible to obtain reliable results below $T/t=0.2$ 
using DQMC. In contrast, 
the MCMF approach easily captures the long-range AFM order as 
shown in \ref{rf-12}(d).

\textit{2. Ground state properties:} 
\begin{figure}[t]
\centering{
\includegraphics[width=8.5cm, height=8.5cm, clip=true]{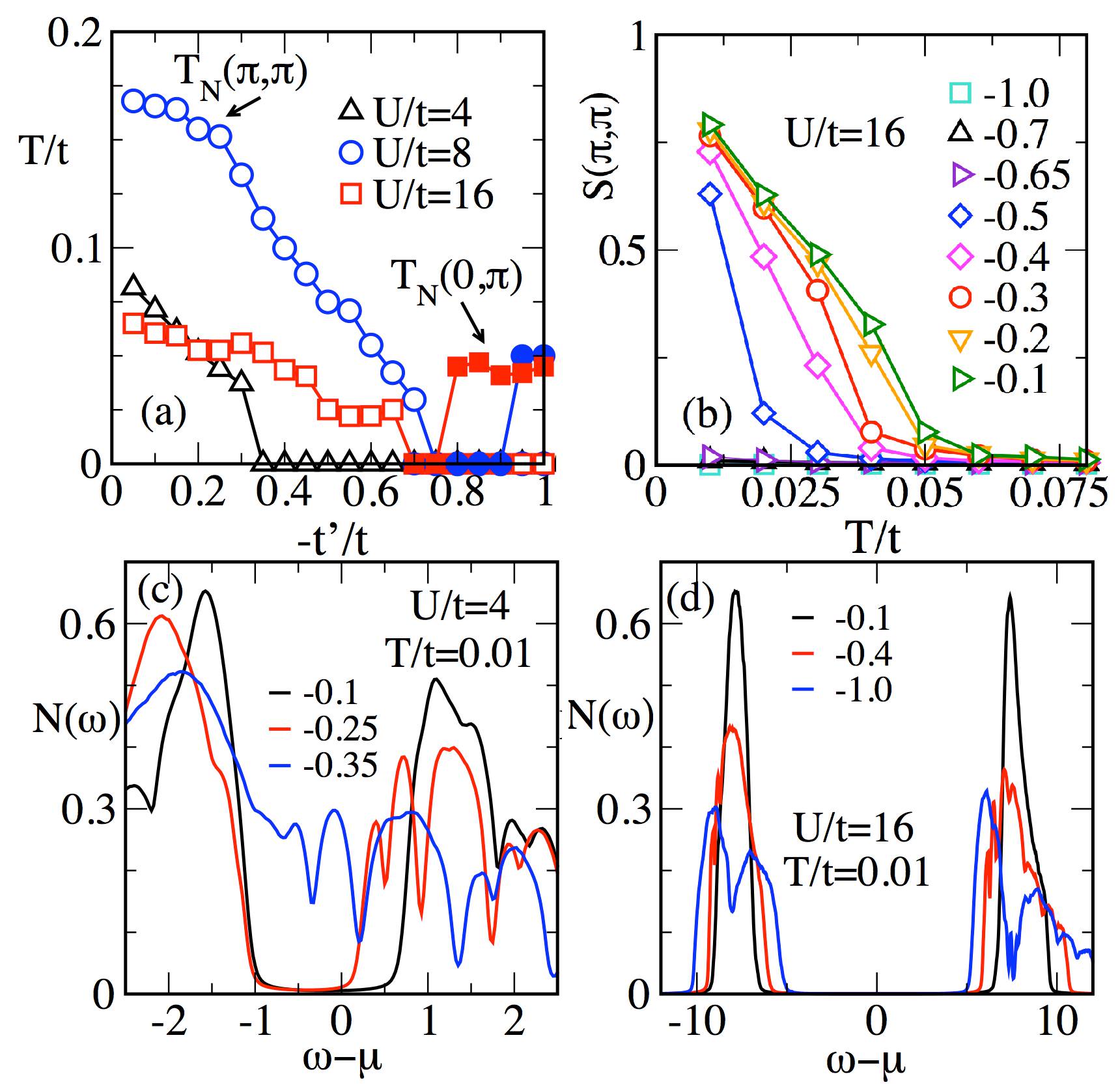}}
\caption{(color online) MCMF results for $t^\prime\neq 0$. Panel (a) 
contains the evolution of $T_N$ with increasing $t^{\prime}/t$ for 
different values of $U/t$ and 
for $\textbf{q}=(\pi,\pi)$ and 
$\textbf{q}=(0,\pi)$. The case $\textbf{q}=(\pi,0)$ is identical 
to $\textbf{q}=(0,\pi)$ and it is not shown. Panel (b) displays 
the typical $S(\pi,\pi)$ for $U/t=16$ at various $t^\prime/t$, as indicated. 
(c) and (d) show $N(\omega)$ for $U/t=4$ and $U/t=16$ respectively, 
at different values of $t^\prime/t$. }
\vspace{-0.0cm} 
\label{rf-13}
\end{figure}
Earlier $T=0$ studies\cite{gs-1,gs-2,gs-3} have established that at small $U$, 
a small finite $t^\prime$ destroys magnetic order in favor of paramagnetism (PM). 
For $t^\prime/t$ below 0.7, the paramagnetic phase evolves into 
a $\textbf{q}=(\pi,\pi)$ antiferromagnet with increasing $U$. 
At larger $t^\prime$ and larger $U$, there is a transition to a state that 
is a linear superposition of $\textbf{q}=(0,\pi)$ and $\textbf{q}=(\pi,0)$ states 
from the PM state. Finally for $U$ greater than $10t$, there is a possible spin 
liquid phase in between the $\textbf{q}=(\pi,\pi)$ and $\textbf{q}=(0,\pi)$/$\textbf{q}=(\pi,0)$ phases. 
The Gutzwiller approximation combined with the random phase approximation (GA+RPA) also find 
a number of incommensurate magnetic phases sandwiched between the low $U$ (PM) and large $U$  
($\textbf{q}=(\pi,\pi)$ or $\textbf{q}=(0,\pi)$/$\textbf{q}=(\pi,0)$) orders.\cite{gs-3}

Here, we present some of the ground state and finite temperature properties 
with the goal to show the ability of MCMF to capture essential physics 
both at low and high temperatures. 
Detailed quantitative comparison with 
existing literature will be presented elsewhere. 
Figure~\ref{rf-13} shows our results. In \ref{rf-13}(a), the locus 
of the $\textbf{q}=(\pi,\pi)$ and $\textbf{q}=(0,\pi)$ N\'eel temperatures 
is shown as a function of $t^\prime/t$. The $U/t$ values used represent small, 
intermediate, and large $U/t$ regimes. At $U/t=4$, the  $\textbf{q}=(\pi,\pi)$ phase 
is progressively weakened and ultimately destroyed in favor of a paramagnetic state. 
In \ref{rf-13}(c) we show the DOS for $U/t=4$. At $t^\prime/t\sim -0.35$, there is an 
insulator to metal transition accompanying the magnetic to PM transition.  
For $U/t=8$ and 16, there is a similar loss of the $\textbf{q}=(\pi,\pi)$ 
magnetic order with increasing $t^\prime/t$. The critical $t^\prime/t$ needed shows non-monotonic dependence on $U/t$ similar to that of $T_N$ with varying $U/t$. The collapse of the $\textbf{q}=(\pi,\pi)$ order with increasing $t^\prime/t$ is shown in Fig. \ref{rf-13}(b) for $U/t=16$. 

In Fig. \ref{rf-13}(d),  the DOS for $U/t=16$ is presented. Clearly, 
the gap in N($\omega$) changes only slightly with $t^\prime/t$ varying from 0 to -1. 
A similar evolution occurs for $U/t=8$ as well. The gap survives because there 
is a transition from $\textbf{q}=(\pi,\pi)$ to a linear combination of the 
$\textbf{q}=(0,\pi) $/$\textbf{q}=(\pi,0)$. The locus of the peak of $\textbf{q}=(0,\pi)$ 
is shown in \ref{rf-13}(a). In the region in between the two phases we only find a weak order 
difficult to distinguish from a PM state. Note that since 
this method reduces to the HF theory at $T=0$, a spin liquid phase cannot be captured 
within this approach due to the lack of quantum fluctuations; however,  
the MCMF method is able to suggest regions in parameter space
where spin liquid phases are possible. 

\section{Conclusions}
In this publication, a many-body technique which is ``intermediate'' between the canonical mean field Hartree-Fock approximation and the numerically exact determinant quantum Monte Carlo method has been discussed and tested for the case of the one-band Hubbard model. The thermal fluctuations that are properly considered in this method were shown to be sufficient to reproduce the expected ``up and down'' non-monotonic behavior of the N\'eel temperature with increasing $U/t$ at half-filling, unveiling a normal state regime where there are preformed local moments but no magnetic long-range order. 

A necessary condition for the new method to work properly is that the mean field approximation used (either the HF method employed here or some other mean field method) captures the essence of the ground state magnetic, orbital, or even superconducting properties. After that step, the MC-MF technique is expected to address reasonably well the temperature fluctuations and generation of short-range order near the critical temperature. There are no obvious restrictions in parameters such as couplings: if the mean field method works at a particular coupling, the MC-MF will work as well varying the temperature. The coupling range where the method works best, in the sense of improving substantially over naive mean-field finite-T approximations, is strong coupling where fluctuations start developing when cooling down at temperatures much higher than the true long-range order critical temperature. Our technique captures the regime where local moments are formed but they are coupled effectively only at short distances. In superconducting systems with strong attraction, the method would capture the formation of individual Cooper pairs upon cooling, followed at lower temperatures by the true superconducting state.

Another advantage of the MCMF method is that it can be applied to 
other Hubbard models that cannot be treated by DQMC due to the 
fermion sign problem.  As in the case of the addition of realistic $t’<0$ next nearest neighbor hopping amplitudes where DQMC can not reach the ordering temperature upon cooling because of the sign problem, the good performance of the MC-MF remains unchanged with regards to the case t’=0. Thus, examples where the MC-MF approach can be applied include the one-orbital Hubbard model with hopping beyond nearest neighbors, as demonstrated here, or the multi-orbital Hubbard models that are widely discussed for iron-based superconductors. The latter will be the focus of future efforts in this context.

\begin{acknowledgments} 
A. Mukherjee and N.P. were partially supported by the
National Science Foundation under Grant No. DMR-1404375.
S.D. was supported in part by NSFC (11274060).
A. Moreo and E.D. were supported by 
the U.S. Department of Energy, Office of Basic Energy 
Sciences, Materials Sciences and Engineering Division.
\end{acknowledgments} 

\bibliography{af-mc.bib}

\begin{thebibliography}{63}%
\makeatletter
\providecommand \@ifxundefined [1]{%
 \@ifx{#1\undefined}
}%
\providecommand \@ifnum [1]{%
 \ifnum #1\expandafter \@firstoftwo
 \else \expandafter \@secondoftwo
 \fi
}%
\providecommand \@ifx [1]{%
 \ifx #1\expandafter \@firstoftwo
 \else \expandafter \@secondoftwo
 \fi
}%
\providecommand \natexlab [1]{#1}%
\providecommand \enquote  [1]{``#1''}%
\providecommand \bibnamefont  [1]{#1}%
\providecommand \bibfnamefont [1]{#1}%
\providecommand \citenamefont [1]{#1}%
\providecommand \href@noop [0]{\@secondoftwo}%
\providecommand \href [0]{\begingroup \@sanitize@url \@href}%
\providecommand \@href[1]{\@@startlink{#1}\@@href}%
\providecommand \@@href[1]{\endgroup#1\@@endlink}%
\providecommand \@sanitize@url [0]{\catcode `\\12\catcode `\$12\catcode
  `\&12\catcode `\#12\catcode `\^12\catcode `\_12\catcode `\%12\relax}%
\providecommand \@@startlink[1]{}%
\providecommand \@@endlink[0]{}%
\providecommand \url  [0]{\begingroup\@sanitize@url \@url }%
\providecommand \@url [1]{\endgroup\@href {#1}{\urlprefix }}%
\providecommand \urlprefix  [0]{URL }%
\providecommand \Eprint [0]{\href }%
\providecommand \doibase [0]{http://dx.doi.org/}%
\providecommand \selectlanguage [0]{\@gobble}%
\providecommand \bibinfo  [0]{\@secondoftwo}%
\providecommand \bibfield  [0]{\@secondoftwo}%
\providecommand \translation [1]{[#1]}%
\providecommand \BibitemOpen [0]{}%
\providecommand \bibitemStop [0]{}%
\providecommand \bibitemNoStop [0]{.\EOS\space}%
\providecommand \EOS [0]{\spacefactor3000\relax}%
\providecommand \BibitemShut  [1]{\csname bibitem#1\endcsname}%
\let\auto@bib@innerbib\@empty
\bibitem [{\citenamefont {Tokura}\ and\ \citenamefont
  {Nagaosa}(2000)}]{tokura}%
  \BibitemOpen
  \bibfield  {author} {\bibinfo {author} {\bibfnamefont {Y.}~\bibnamefont
  {Tokura}}\ and\ \bibinfo {author} {\bibfnamefont {N.}~\bibnamefont
  {Nagaosa}},\ }\href {\doibase 10.1126/science.288.5465.462} {\bibfield
  {journal} {\bibinfo  {journal} {Science}\ }\textbf {\bibinfo {volume}
  {288}},\ \bibinfo {pages} {462} (\bibinfo {year} {2000})}\BibitemShut
  {NoStop}%
\bibitem [{\citenamefont {Dagotto}(2005)}]{complex}%
  \BibitemOpen
  \bibfield  {author} {\bibinfo {author} {\bibfnamefont {E.}~\bibnamefont
  {Dagotto}},\ }\href {\doibase 10.1126/science.1107559} {\bibfield  {journal}
  {\bibinfo  {journal} {Science}\ }\textbf {\bibinfo {volume} {309}},\ \bibinfo
  {pages} {257} (\bibinfo {year} {2005})}\BibitemShut {NoStop}%
\bibitem [{\citenamefont {Dagotto}(1994)}]{RMP94}%
  \BibitemOpen
  \bibfield  {author} {\bibinfo {author} {\bibfnamefont {E.}~\bibnamefont
  {Dagotto}},\ }\href {\doibase 10.1103/RevModPhys.66.763} {\bibfield
  {journal} {\bibinfo  {journal} {Rev. Mod. Phys.}\ }\textbf {\bibinfo {volume}
  {66}},\ \bibinfo {pages} {763} (\bibinfo {year} {1994})}\BibitemShut
  {NoStop}%
\bibitem [{\citenamefont {Scalapino}(1995)}]{scalapino}%
  \BibitemOpen
  \bibfield  {author} {\bibinfo {author} {\bibfnamefont {D.}~\bibnamefont
  {Scalapino}},\ }\href {\doibase
  http://dx.doi.org/10.1016/0370-1573(94)00086-I} {\bibfield  {journal}
  {\bibinfo  {journal} {Physics Reports}\ }\textbf {\bibinfo {volume} {250}},\
  \bibinfo {pages} {329 } (\bibinfo {year} {1995})}\BibitemShut {NoStop}%
\bibitem [{\citenamefont {Lee}\ \emph {et~al.}(2006)\citenamefont {Lee},
  \citenamefont {Nagaosa},\ and\ \citenamefont {Wen}}]{lee}%
  \BibitemOpen
  \bibfield  {author} {\bibinfo {author} {\bibfnamefont {P.~A.}\ \bibnamefont
  {Lee}}, \bibinfo {author} {\bibfnamefont {N.}~\bibnamefont {Nagaosa}}, \ and\
  \bibinfo {author} {\bibfnamefont {X.-G.}\ \bibnamefont {Wen}},\ }\href
  {\doibase 10.1103/RevModPhys.78.17} {\bibfield  {journal} {\bibinfo
  {journal} {Rev. Mod. Phys.}\ }\textbf {\bibinfo {volume} {78}},\ \bibinfo
  {pages} {17} (\bibinfo {year} {2006})}\BibitemShut {NoStop}%
\bibitem [{\citenamefont {White}(1992)}]{dmrg}%
  \BibitemOpen
  \bibfield  {author} {\bibinfo {author} {\bibfnamefont {S.~R.}\ \bibnamefont
  {White}},\ }\href {\doibase 10.1103/PhysRevLett.69.2863} {\bibfield
  {journal} {\bibinfo  {journal} {Phys. Rev. Lett.}\ }\textbf {\bibinfo
  {volume} {69}},\ \bibinfo {pages} {2863} (\bibinfo {year}
  {1992})}\BibitemShut {NoStop}%
\bibitem [{\citenamefont {Blankenbecler}\ \emph {et~al.}(1981)\citenamefont
  {Blankenbecler}, \citenamefont {Scalapino},\ and\ \citenamefont
  {Sugar}}]{dqmc}%
  \BibitemOpen
  \bibfield  {author} {\bibinfo {author} {\bibfnamefont {R.}~\bibnamefont
  {Blankenbecler}}, \bibinfo {author} {\bibfnamefont {D.~J.}\ \bibnamefont
  {Scalapino}}, \ and\ \bibinfo {author} {\bibfnamefont {R.~L.}\ \bibnamefont
  {Sugar}},\ }\href {\doibase 10.1103/PhysRevD.24.2278} {\bibfield  {journal}
  {\bibinfo  {journal} {Phys. Rev. D}\ }\textbf {\bibinfo {volume} {24}},\
  \bibinfo {pages} {2278} (\bibinfo {year} {1981})}\BibitemShut {NoStop}%
\bibitem [{\citenamefont {White}\ \emph {et~al.}(1989)\citenamefont {White},
  \citenamefont {Scalapino}, \citenamefont {Sugar}, \citenamefont {Loh},
  \citenamefont {Gubernatis},\ and\ \citenamefont {Scalettar}}]{WhitePRB1989}%
  \BibitemOpen
  \bibfield  {author} {\bibinfo {author} {\bibfnamefont {S.~R.}\ \bibnamefont
  {White}}, \bibinfo {author} {\bibfnamefont {D.~J.}\ \bibnamefont
  {Scalapino}}, \bibinfo {author} {\bibfnamefont {R.~L.}\ \bibnamefont
  {Sugar}}, \bibinfo {author} {\bibfnamefont {E.~Y.}\ \bibnamefont {Loh}},
  \bibinfo {author} {\bibfnamefont {J.~E.}\ \bibnamefont {Gubernatis}}, \ and\
  \bibinfo {author} {\bibfnamefont {R.~T.}\ \bibnamefont {Scalettar}},\ }\href
  {\doibase 10.1103/PhysRevB.40.506} {\bibfield  {journal} {\bibinfo  {journal}
  {Phys. Rev. B}\ }\textbf {\bibinfo {volume} {40}},\ \bibinfo {pages} {506}
  (\bibinfo {year} {1989})}\BibitemShut {NoStop}%
\bibitem [{\citenamefont {Paiva}\ \emph {et~al.}(2010)\citenamefont {Paiva},
  \citenamefont {Scalettar}, \citenamefont {Randeria},\ and\ \citenamefont
  {Trivedi}}]{paiva10}%
  \BibitemOpen
  \bibfield  {author} {\bibinfo {author} {\bibfnamefont {T.}~\bibnamefont
  {Paiva}}, \bibinfo {author} {\bibfnamefont {R.}~\bibnamefont {Scalettar}},
  \bibinfo {author} {\bibfnamefont {M.}~\bibnamefont {Randeria}}, \ and\
  \bibinfo {author} {\bibfnamefont {N.}~\bibnamefont {Trivedi}},\ }\href
  {\doibase 10.1103/PhysRevLett.104.066406} {\bibfield  {journal} {\bibinfo
  {journal} {Phys. Rev. Lett.}\ }\textbf {\bibinfo {volume} {104}},\ \bibinfo
  {pages} {066406} (\bibinfo {year} {2010})},\ \bibinfo {note} {and references
  therein.}\BibitemShut {Stop}%
\bibitem [{\citenamefont {Bloch}\ \emph {et~al.}(2008)\citenamefont {Bloch},
  \citenamefont {Dalibard},\ and\ \citenamefont {Zwerger}}]{optical-lattices}%
  \BibitemOpen
  \bibfield  {author} {\bibinfo {author} {\bibfnamefont {I.}~\bibnamefont
  {Bloch}}, \bibinfo {author} {\bibfnamefont {J.}~\bibnamefont {Dalibard}}, \
  and\ \bibinfo {author} {\bibfnamefont {W.}~\bibnamefont {Zwerger}},\ }\href
  {\doibase 10.1103/RevModPhys.80.885} {\bibfield  {journal} {\bibinfo
  {journal} {Rev. Mod. Phys.}\ }\textbf {\bibinfo {volume} {80}},\ \bibinfo
  {pages} {885} (\bibinfo {year} {2008})}\BibitemShut {NoStop}%
\bibitem [{\citenamefont {Paiva}\ \emph {et~al.}(2011)\citenamefont {Paiva},
  \citenamefont {Loh}, \citenamefont {Randeria}, \citenamefont {Scalettar},\
  and\ \citenamefont {Trivedi}}]{paiva}%
  \BibitemOpen
  \bibfield  {author} {\bibinfo {author} {\bibfnamefont {T.}~\bibnamefont
  {Paiva}}, \bibinfo {author} {\bibfnamefont {Y.~L.}\ \bibnamefont {Loh}},
  \bibinfo {author} {\bibfnamefont {M.}~\bibnamefont {Randeria}}, \bibinfo
  {author} {\bibfnamefont {R.~T.}\ \bibnamefont {Scalettar}}, \ and\ \bibinfo
  {author} {\bibfnamefont {N.}~\bibnamefont {Trivedi}},\ }\href {\doibase
  10.1103/PhysRevLett.107.086401} {\bibfield  {journal} {\bibinfo  {journal}
  {Phys. Rev. Lett.}\ }\textbf {\bibinfo {volume} {107}},\ \bibinfo {pages}
  {086401} (\bibinfo {year} {2011})}\BibitemShut {NoStop}%
\bibitem [{\citenamefont {Kozik}\ \emph {et~al.}(2013)\citenamefont {Kozik},
  \citenamefont {Burovski}, \citenamefont {Scarola},\ and\ \citenamefont
  {Troyer}}]{kozik}%
  \BibitemOpen
  \bibfield  {author} {\bibinfo {author} {\bibfnamefont {E.}~\bibnamefont
  {Kozik}}, \bibinfo {author} {\bibfnamefont {E.}~\bibnamefont {Burovski}},
  \bibinfo {author} {\bibfnamefont {V.~W.}\ \bibnamefont {Scarola}}, \ and\
  \bibinfo {author} {\bibfnamefont {M.}~\bibnamefont {Troyer}},\ }\href
  {\doibase 10.1103/PhysRevB.87.205102} {\bibfield  {journal} {\bibinfo
  {journal} {Phys. Rev. B}\ }\textbf {\bibinfo {volume} {87}},\ \bibinfo
  {pages} {205102} (\bibinfo {year} {2013})}\BibitemShut {NoStop}%
\bibitem [{\citenamefont {Loh}\ \emph {et~al.}(1990)\citenamefont {Loh},
  \citenamefont {Gubernatis}, \citenamefont {Scalettar}, \citenamefont {White},
  \citenamefont {Scalapino},\ and\ \citenamefont {Sugar}}]{sign}%
  \BibitemOpen
  \bibfield  {author} {\bibinfo {author} {\bibfnamefont {E.~Y.}\ \bibnamefont
  {Loh}}, \bibinfo {author} {\bibfnamefont {J.~E.}\ \bibnamefont {Gubernatis}},
  \bibinfo {author} {\bibfnamefont {R.~T.}\ \bibnamefont {Scalettar}}, \bibinfo
  {author} {\bibfnamefont {S.~R.}\ \bibnamefont {White}}, \bibinfo {author}
  {\bibfnamefont {D.~J.}\ \bibnamefont {Scalapino}}, \ and\ \bibinfo {author}
  {\bibfnamefont {R.~L.}\ \bibnamefont {Sugar}},\ }\href {\doibase
  10.1103/PhysRevB.41.9301} {\bibfield  {journal} {\bibinfo  {journal} {Phys.
  Rev. B}\ }\textbf {\bibinfo {volume} {41}},\ \bibinfo {pages} {9301}
  (\bibinfo {year} {1990})}\BibitemShut {NoStop}%
\bibitem [{\citenamefont {Johnston}(2010)}]{johnston}%
  \BibitemOpen
  \bibfield  {author} {\bibinfo {author} {\bibfnamefont {D.~C.}\ \bibnamefont
  {Johnston}},\ }\href {\doibase 10.1080/00018732.2010.513480} {\bibfield
  {journal} {\bibinfo  {journal} {Adv. Phys.}\ }\textbf {\bibinfo {volume}
  {59}},\ \bibinfo {pages} {803} (\bibinfo {year} {2010})}\BibitemShut
  {NoStop}%
\bibitem [{\citenamefont {Stewart}(2011)}]{stewart}%
  \BibitemOpen
  \bibfield  {author} {\bibinfo {author} {\bibfnamefont {G.~R.}\ \bibnamefont
  {Stewart}},\ }\href {\doibase 10.1103/RevModPhys.83.1589} {\bibfield
  {journal} {\bibinfo  {journal} {Rev. Mod. Phys.}\ }\textbf {\bibinfo {volume}
  {83}},\ \bibinfo {pages} {1589} (\bibinfo {year} {2011})}\BibitemShut
  {NoStop}%
\bibitem [{\citenamefont {Dai}\ \emph {et~al.}(2012)\citenamefont {Dai},
  \citenamefont {Hu},\ and\ \citenamefont {Dagotto}}]{natphys12}%
  \BibitemOpen
  \bibfield  {author} {\bibinfo {author} {\bibfnamefont {P.}~\bibnamefont
  {Dai}}, \bibinfo {author} {\bibfnamefont {J.}~\bibnamefont {Hu}}, \ and\
  \bibinfo {author} {\bibfnamefont {E.}~\bibnamefont {Dagotto}},\ }\href
  {http://dx.doi.org/10.1038/nphys2438} {\bibfield  {journal} {\bibinfo
  {journal} {Nat Phys}\ }\textbf {\bibinfo {volume} {8}},\ \bibinfo {pages}
  {709} (\bibinfo {year} {2012})}\BibitemShut {NoStop}%
\bibitem [{\citenamefont {Dagotto}(2013)}]{RMP13}%
  \BibitemOpen
  \bibfield  {author} {\bibinfo {author} {\bibfnamefont {E.}~\bibnamefont
  {Dagotto}},\ }\href {\doibase 10.1103/RevModPhys.85.849} {\bibfield
  {journal} {\bibinfo  {journal} {Rev. Mod. Phys.}\ }\textbf {\bibinfo {volume}
  {85}},\ \bibinfo {pages} {849} (\bibinfo {year} {2013})}\BibitemShut
  {NoStop}%
\bibitem [{\citenamefont {Hirschfeld}\ \emph {et~al.}(2011)\citenamefont
  {Hirschfeld}, \citenamefont {Korshunov},\ and\ \citenamefont
  {Mazin}}]{peter}%
  \BibitemOpen
  \bibfield  {author} {\bibinfo {author} {\bibfnamefont {P.~J.}\ \bibnamefont
  {Hirschfeld}}, \bibinfo {author} {\bibfnamefont {M.~M.}\ \bibnamefont
  {Korshunov}}, \ and\ \bibinfo {author} {\bibfnamefont {I.~I.}\ \bibnamefont
  {Mazin}},\ }\href {http://stacks.iop.org/0034-4885/74/i=12/a=124508}
  {\bibfield  {journal} {\bibinfo  {journal} {Reports on Progress in Physics}\
  }\textbf {\bibinfo {volume} {74}},\ \bibinfo {pages} {124508} (\bibinfo
  {year} {2011})}\BibitemShut {NoStop}%
\bibitem [{\citenamefont {Si}\ and\ \citenamefont {Abrahams}(2008)}]{si}%
  \BibitemOpen
  \bibfield  {author} {\bibinfo {author} {\bibfnamefont {Q.}~\bibnamefont
  {Si}}\ and\ \bibinfo {author} {\bibfnamefont {E.}~\bibnamefont {Abrahams}},\
  }\href {\doibase 10.1103/PhysRevLett.101.076401} {\bibfield  {journal}
  {\bibinfo  {journal} {Phys. Rev. Lett.}\ }\textbf {\bibinfo {volume} {101}},\
  \bibinfo {pages} {076401} (\bibinfo {year} {2008})}\BibitemShut {NoStop}%
\bibitem [{\citenamefont {Daghofer}\ \emph {et~al.}(2010)\citenamefont
  {Daghofer}, \citenamefont {Nicholson}, \citenamefont {Moreo},\ and\
  \citenamefont {Dagotto}}]{three}%
  \BibitemOpen
  \bibfield  {author} {\bibinfo {author} {\bibfnamefont {M.}~\bibnamefont
  {Daghofer}}, \bibinfo {author} {\bibfnamefont {A.}~\bibnamefont {Nicholson}},
  \bibinfo {author} {\bibfnamefont {A.}~\bibnamefont {Moreo}}, \ and\ \bibinfo
  {author} {\bibfnamefont {E.}~\bibnamefont {Dagotto}},\ }\href {\doibase
  10.1103/PhysRevB.81.014511} {\bibfield  {journal} {\bibinfo  {journal} {Phys.
  Rev. B}\ }\textbf {\bibinfo {volume} {81}},\ \bibinfo {pages} {014511}
  (\bibinfo {year} {2010})},\ \bibinfo {note} {and references
  therein.}\BibitemShut {Stop}%
\bibitem [{\citenamefont {Dagotto}\ \emph {et~al.}(2001)\citenamefont
  {Dagotto}, \citenamefont {Hotta},\ and\ \citenamefont {Moreo}}]{physrep}%
  \BibitemOpen
  \bibfield  {author} {\bibinfo {author} {\bibfnamefont {E.}~\bibnamefont
  {Dagotto}}, \bibinfo {author} {\bibfnamefont {T.}~\bibnamefont {Hotta}}, \
  and\ \bibinfo {author} {\bibfnamefont {A.}~\bibnamefont {Moreo}},\ }\href
  {\doibase http://dx.doi.org/10.1016/S0370-1573(00)00121-6} {\bibfield
  {journal} {\bibinfo  {journal} {Physics Reports}\ }\textbf {\bibinfo {volume}
  {344}},\ \bibinfo {pages} {1 } (\bibinfo {year} {2001})}\BibitemShut
  {NoStop}%
\bibitem [{\citenamefont {Salamon}\ and\ \citenamefont
  {Jaime}(2001)}]{salamon}%
  \BibitemOpen
  \bibfield  {author} {\bibinfo {author} {\bibfnamefont {M.~B.}\ \bibnamefont
  {Salamon}}\ and\ \bibinfo {author} {\bibfnamefont {M.}~\bibnamefont
  {Jaime}},\ }\href {\doibase 10.1103/RevModPhys.73.583} {\bibfield  {journal}
  {\bibinfo  {journal} {Rev. Mod. Phys.}\ }\textbf {\bibinfo {volume} {73}},\
  \bibinfo {pages} {583} (\bibinfo {year} {2001})}\BibitemShut {NoStop}%
\bibitem [{\citenamefont {Dagotto}\ \emph {et~al.}(1998)\citenamefont
  {Dagotto}, \citenamefont {Yunoki}, \citenamefont {Malvezzi}, \citenamefont
  {Moreo}, \citenamefont {Hu}, \citenamefont {Capponi}, \citenamefont
  {Poilblanc},\ and\ \citenamefont {Furukawa}}]{dmrgmanga}%
  \BibitemOpen
  \bibfield  {author} {\bibinfo {author} {\bibfnamefont {E.}~\bibnamefont
  {Dagotto}}, \bibinfo {author} {\bibfnamefont {S.}~\bibnamefont {Yunoki}},
  \bibinfo {author} {\bibfnamefont {A.~L.}\ \bibnamefont {Malvezzi}}, \bibinfo
  {author} {\bibfnamefont {A.}~\bibnamefont {Moreo}}, \bibinfo {author}
  {\bibfnamefont {J.}~\bibnamefont {Hu}}, \bibinfo {author} {\bibfnamefont
  {S.}~\bibnamefont {Capponi}}, \bibinfo {author} {\bibfnamefont
  {D.}~\bibnamefont {Poilblanc}}, \ and\ \bibinfo {author} {\bibfnamefont
  {N.}~\bibnamefont {Furukawa}},\ }\href {\doibase 10.1103/PhysRevB.58.6414}
  {\bibfield  {journal} {\bibinfo  {journal} {Phys. Rev. B}\ }\textbf {\bibinfo
  {volume} {58}},\ \bibinfo {pages} {6414} (\bibinfo {year}
  {1998})}\BibitemShut {NoStop}%
\bibitem [{\citenamefont {Yunoki}\ \emph {et~al.}(1998)\citenamefont {Yunoki},
  \citenamefont {Hu}, \citenamefont {Malvezzi}, \citenamefont {Moreo},
  \citenamefont {Furukawa},\ and\ \citenamefont {Dagotto}}]{yunoki}%
  \BibitemOpen
  \bibfield  {author} {\bibinfo {author} {\bibfnamefont {S.}~\bibnamefont
  {Yunoki}}, \bibinfo {author} {\bibfnamefont {J.}~\bibnamefont {Hu}}, \bibinfo
  {author} {\bibfnamefont {A.~L.}\ \bibnamefont {Malvezzi}}, \bibinfo {author}
  {\bibfnamefont {A.}~\bibnamefont {Moreo}}, \bibinfo {author} {\bibfnamefont
  {N.}~\bibnamefont {Furukawa}}, \ and\ \bibinfo {author} {\bibfnamefont
  {E.}~\bibnamefont {Dagotto}},\ }\href {\doibase 10.1103/PhysRevLett.80.845}
  {\bibfield  {journal} {\bibinfo  {journal} {Phys. Rev. Lett.}\ }\textbf
  {\bibinfo {volume} {80}},\ \bibinfo {pages} {845} (\bibinfo {year}
  {1998})}\BibitemShut {NoStop}%
\bibitem [{\citenamefont {Moreo}\ \emph {et~al.}(1999)\citenamefont {Moreo},
  \citenamefont {Yunoki},\ and\ \citenamefont {Dagotto}}]{science99}%
  \BibitemOpen
  \bibfield  {author} {\bibinfo {author} {\bibfnamefont {A.}~\bibnamefont
  {Moreo}}, \bibinfo {author} {\bibfnamefont {S.}~\bibnamefont {Yunoki}}, \
  and\ \bibinfo {author} {\bibfnamefont {E.}~\bibnamefont {Dagotto}},\ }\href
  {\doibase 10.1126/science.283.5410.2034} {\bibfield  {journal} {\bibinfo
  {journal} {Science}\ }\textbf {\bibinfo {volume} {283}},\ \bibinfo {pages}
  {2034} (\bibinfo {year} {1999})}\BibitemShut {NoStop}%
\bibitem [{\citenamefont {Kumar}\ and\ \citenamefont
  {Majumdar}(2006{\natexlab{a}})}]{kumar}%
  \BibitemOpen
  \bibfield  {author} {\bibinfo {author} {\bibfnamefont {S.}~\bibnamefont
  {Kumar}}\ and\ \bibinfo {author} {\bibfnamefont {P.}~\bibnamefont
  {Majumdar}},\ }\href {\doibase 10.1103/PhysRevLett.96.016602} {\bibfield
  {journal} {\bibinfo  {journal} {Phys. Rev. Lett.}\ }\textbf {\bibinfo
  {volume} {96}},\ \bibinfo {pages} {016602} (\bibinfo {year}
  {2006}{\natexlab{a}})}\BibitemShut {NoStop}%
\bibitem [{\citenamefont {Dong}\ \emph {et~al.}(2008)\citenamefont {Dong},
  \citenamefont {Yu}, \citenamefont {Yunoki}, \citenamefont {Liu},\ and\
  \citenamefont {Dagotto}}]{dong1}%
  \BibitemOpen
  \bibfield  {author} {\bibinfo {author} {\bibfnamefont {S.}~\bibnamefont
  {Dong}}, \bibinfo {author} {\bibfnamefont {R.}~\bibnamefont {Yu}}, \bibinfo
  {author} {\bibfnamefont {S.}~\bibnamefont {Yunoki}}, \bibinfo {author}
  {\bibfnamefont {J.-M.}\ \bibnamefont {Liu}}, \ and\ \bibinfo {author}
  {\bibfnamefont {E.}~\bibnamefont {Dagotto}},\ }\href {\doibase
  10.1103/PhysRevB.78.155121} {\bibfield  {journal} {\bibinfo  {journal} {Phys.
  Rev. B}\ }\textbf {\bibinfo {volume} {78}},\ \bibinfo {pages} {155121}
  (\bibinfo {year} {2008})}\BibitemShut {NoStop}%
\bibitem [{\citenamefont {Dong}\ \emph {et~al.}(2009)\citenamefont {Dong},
  \citenamefont {Yu}, \citenamefont {Liu},\ and\ \citenamefont
  {Dagotto}}]{dong2}%
  \BibitemOpen
  \bibfield  {author} {\bibinfo {author} {\bibfnamefont {S.}~\bibnamefont
  {Dong}}, \bibinfo {author} {\bibfnamefont {R.}~\bibnamefont {Yu}}, \bibinfo
  {author} {\bibfnamefont {J.-M.}\ \bibnamefont {Liu}}, \ and\ \bibinfo
  {author} {\bibfnamefont {E.}~\bibnamefont {Dagotto}},\ }\href {\doibase
  10.1103/PhysRevLett.103.107204} {\bibfield  {journal} {\bibinfo  {journal}
  {Phys. Rev. Lett.}\ }\textbf {\bibinfo {volume} {103}},\ \bibinfo {pages}
  {107204} (\bibinfo {year} {2009})}\BibitemShut {NoStop}%
\bibitem [{\citenamefont {Liang}\ \emph {et~al.}(2011)\citenamefont {Liang},
  \citenamefont {Daghofer}, \citenamefont {Dong}, \citenamefont
  {\ifmmode~\mbox{\c{S}}\else \c{S}\fi{}en},\ and\ \citenamefont
  {Dagotto}}]{liang11}%
  \BibitemOpen
  \bibfield  {author} {\bibinfo {author} {\bibfnamefont {S.}~\bibnamefont
  {Liang}}, \bibinfo {author} {\bibfnamefont {M.}~\bibnamefont {Daghofer}},
  \bibinfo {author} {\bibfnamefont {S.}~\bibnamefont {Dong}}, \bibinfo {author}
  {\bibfnamefont {C.}~\bibnamefont {\ifmmode~\mbox{\c{S}}\else \c{S}\fi{}en}},
  \ and\ \bibinfo {author} {\bibfnamefont {E.}~\bibnamefont {Dagotto}},\ }\href
  {\doibase 10.1103/PhysRevB.84.024408} {\bibfield  {journal} {\bibinfo
  {journal} {Phys. Rev. B}\ }\textbf {\bibinfo {volume} {84}},\ \bibinfo
  {pages} {024408} (\bibinfo {year} {2011})}\BibitemShut {NoStop}%
\bibitem [{\citenamefont {\ifmmode~\mbox{\c{S}}\else \c{S}\fi{}en}\ \emph
  {et~al.}(2012)\citenamefont {\ifmmode~\mbox{\c{S}}\else \c{S}\fi{}en},
  \citenamefont {Liang},\ and\ \citenamefont {Dagotto}}]{sen}%
  \BibitemOpen
  \bibfield  {author} {\bibinfo {author} {\bibfnamefont {C.}~\bibnamefont
  {\ifmmode~\mbox{\c{S}}\else \c{S}\fi{}en}}, \bibinfo {author} {\bibfnamefont
  {S.}~\bibnamefont {Liang}}, \ and\ \bibinfo {author} {\bibfnamefont
  {E.}~\bibnamefont {Dagotto}},\ }\href {\doibase 10.1103/PhysRevB.85.174418}
  {\bibfield  {journal} {\bibinfo  {journal} {Phys. Rev. B}\ }\textbf {\bibinfo
  {volume} {85}},\ \bibinfo {pages} {174418} (\bibinfo {year}
  {2012})}\BibitemShut {NoStop}%
\bibitem [{\citenamefont {Buhler}\ \emph
  {et~al.}(2000{\natexlab{a}})\citenamefont {Buhler}, \citenamefont {Yunoki},\
  and\ \citenamefont {Moreo}}]{buhler1}%
  \BibitemOpen
  \bibfield  {author} {\bibinfo {author} {\bibfnamefont {C.}~\bibnamefont
  {Buhler}}, \bibinfo {author} {\bibfnamefont {S.}~\bibnamefont {Yunoki}}, \
  and\ \bibinfo {author} {\bibfnamefont {A.}~\bibnamefont {Moreo}},\ }\href
  {\doibase 10.1103/PhysRevLett.84.2690} {\bibfield  {journal} {\bibinfo
  {journal} {Phys. Rev. Lett.}\ }\textbf {\bibinfo {volume} {84}},\ \bibinfo
  {pages} {2690} (\bibinfo {year} {2000}{\natexlab{a}})}\BibitemShut {NoStop}%
\bibitem [{\citenamefont {Buhler}\ \emph
  {et~al.}(2000{\natexlab{b}})\citenamefont {Buhler}, \citenamefont {Yunoki},\
  and\ \citenamefont {Moreo}}]{buhler2}%
  \BibitemOpen
  \bibfield  {author} {\bibinfo {author} {\bibfnamefont {C.}~\bibnamefont
  {Buhler}}, \bibinfo {author} {\bibfnamefont {S.}~\bibnamefont {Yunoki}}, \
  and\ \bibinfo {author} {\bibfnamefont {A.}~\bibnamefont {Moreo}},\ }\href
  {\doibase 10.1103/PhysRevB.62.R3620} {\bibfield  {journal} {\bibinfo
  {journal} {Phys. Rev. B}\ }\textbf {\bibinfo {volume} {62}},\ \bibinfo
  {pages} {R3620} (\bibinfo {year} {2000}{\natexlab{b}})}\BibitemShut {NoStop}%
\bibitem [{\citenamefont {Moraghebi}\ \emph {et~al.}(2001)\citenamefont
  {Moraghebi}, \citenamefont {Buhler}, \citenamefont {Yunoki},\ and\
  \citenamefont {Moreo}}]{mora1}%
  \BibitemOpen
  \bibfield  {author} {\bibinfo {author} {\bibfnamefont {M.}~\bibnamefont
  {Moraghebi}}, \bibinfo {author} {\bibfnamefont {C.}~\bibnamefont {Buhler}},
  \bibinfo {author} {\bibfnamefont {S.}~\bibnamefont {Yunoki}}, \ and\ \bibinfo
  {author} {\bibfnamefont {A.}~\bibnamefont {Moreo}},\ }\href {\doibase
  10.1103/PhysRevB.63.214513} {\bibfield  {journal} {\bibinfo  {journal} {Phys.
  Rev. B}\ }\textbf {\bibinfo {volume} {63}},\ \bibinfo {pages} {214513}
  (\bibinfo {year} {2001})}\BibitemShut {NoStop}%
\bibitem [{\citenamefont {Moraghebi}\ \emph
  {et~al.}(2002{\natexlab{a}})\citenamefont {Moraghebi}, \citenamefont
  {Yunoki},\ and\ \citenamefont {Moreo}}]{mora2}%
  \BibitemOpen
  \bibfield  {author} {\bibinfo {author} {\bibfnamefont {M.}~\bibnamefont
  {Moraghebi}}, \bibinfo {author} {\bibfnamefont {S.}~\bibnamefont {Yunoki}}, \
  and\ \bibinfo {author} {\bibfnamefont {A.}~\bibnamefont {Moreo}},\ }\href
  {\doibase 10.1103/PhysRevB.66.214522} {\bibfield  {journal} {\bibinfo
  {journal} {Phys. Rev. B}\ }\textbf {\bibinfo {volume} {66}},\ \bibinfo
  {pages} {214522} (\bibinfo {year} {2002}{\natexlab{a}})}\BibitemShut
  {NoStop}%
\bibitem [{\citenamefont {Moraghebi}\ \emph
  {et~al.}(2002{\natexlab{b}})\citenamefont {Moraghebi}, \citenamefont
  {Yunoki},\ and\ \citenamefont {Moreo}}]{mora3}%
  \BibitemOpen
  \bibfield  {author} {\bibinfo {author} {\bibfnamefont {M.}~\bibnamefont
  {Moraghebi}}, \bibinfo {author} {\bibfnamefont {S.}~\bibnamefont {Yunoki}}, \
  and\ \bibinfo {author} {\bibfnamefont {A.}~\bibnamefont {Moreo}},\ }\href
  {\doibase 10.1103/PhysRevLett.88.187001} {\bibfield  {journal} {\bibinfo
  {journal} {Phys. Rev. Lett.}\ }\textbf {\bibinfo {volume} {88}},\ \bibinfo
  {pages} {187001} (\bibinfo {year} {2002}{\natexlab{b}})}\BibitemShut
  {NoStop}%
\bibitem [{\citenamefont {Mayr}\ \emph {et~al.}(2005)\citenamefont {Mayr},
  \citenamefont {Alvarez}, \citenamefont {\ifmmode~\mbox{\c{S}}\else
  \c{S}\fi{}en},\ and\ \citenamefont {Dagotto}}]{BdG0}%
  \BibitemOpen
  \bibfield  {author} {\bibinfo {author} {\bibfnamefont {M.}~\bibnamefont
  {Mayr}}, \bibinfo {author} {\bibfnamefont {G.}~\bibnamefont {Alvarez}},
  \bibinfo {author} {\bibfnamefont {C.}~\bibnamefont
  {\ifmmode~\mbox{\c{S}}\else \c{S}\fi{}en}}, \ and\ \bibinfo {author}
  {\bibfnamefont {E.}~\bibnamefont {Dagotto}},\ }\href {\doibase
  10.1103/PhysRevLett.94.217001} {\bibfield  {journal} {\bibinfo  {journal}
  {Phys. Rev. Lett.}\ }\textbf {\bibinfo {volume} {94}},\ \bibinfo {pages}
  {217001} (\bibinfo {year} {2005})}\BibitemShut {NoStop}%
\bibitem [{\citenamefont {Alvarez}\ \emph {et~al.}(2005)\citenamefont
  {Alvarez}, \citenamefont {Mayr}, \citenamefont {Moreo},\ and\ \citenamefont
  {Dagotto}}]{BdG1}%
  \BibitemOpen
  \bibfield  {author} {\bibinfo {author} {\bibfnamefont {G.}~\bibnamefont
  {Alvarez}}, \bibinfo {author} {\bibfnamefont {M.}~\bibnamefont {Mayr}},
  \bibinfo {author} {\bibfnamefont {A.}~\bibnamefont {Moreo}}, \ and\ \bibinfo
  {author} {\bibfnamefont {E.}~\bibnamefont {Dagotto}},\ }\href {\doibase
  10.1103/PhysRevB.71.014514} {\bibfield  {journal} {\bibinfo  {journal} {Phys.
  Rev. B}\ }\textbf {\bibinfo {volume} {71}},\ \bibinfo {pages} {014514}
  (\bibinfo {year} {2005})}\BibitemShut {NoStop}%
\bibitem [{\citenamefont {Mayr}\ \emph {et~al.}(2006)\citenamefont {Mayr},
  \citenamefont {Alvarez}, \citenamefont {Moreo},\ and\ \citenamefont
  {Dagotto}}]{BdG2}%
  \BibitemOpen
  \bibfield  {author} {\bibinfo {author} {\bibfnamefont {M.}~\bibnamefont
  {Mayr}}, \bibinfo {author} {\bibfnamefont {G.}~\bibnamefont {Alvarez}},
  \bibinfo {author} {\bibfnamefont {A.}~\bibnamefont {Moreo}}, \ and\ \bibinfo
  {author} {\bibfnamefont {E.}~\bibnamefont {Dagotto}},\ }\href {\doibase
  10.1103/PhysRevB.73.014509} {\bibfield  {journal} {\bibinfo  {journal} {Phys.
  Rev. B}\ }\textbf {\bibinfo {volume} {73}},\ \bibinfo {pages} {014509}
  (\bibinfo {year} {2006})}\BibitemShut {NoStop}%
\bibitem [{\citenamefont {Alvarez}\ and\ \citenamefont {Dagotto}(2008)}]{BdG3}%
  \BibitemOpen
  \bibfield  {author} {\bibinfo {author} {\bibfnamefont {G.}~\bibnamefont
  {Alvarez}}\ and\ \bibinfo {author} {\bibfnamefont {E.}~\bibnamefont
  {Dagotto}},\ }\href {\doibase 10.1103/PhysRevLett.101.177001} {\bibfield
  {journal} {\bibinfo  {journal} {Phys. Rev. Lett.}\ }\textbf {\bibinfo
  {volume} {101}},\ \bibinfo {pages} {177001} (\bibinfo {year}
  {2008})}\BibitemShut {NoStop}%
\bibitem [{\citenamefont {Jarrell}\ and\ \citenamefont
  {Gubernatis}(1996)}]{MaxEnt}%
  \BibitemOpen
  \bibfield  {author} {\bibinfo {author} {\bibfnamefont {M.}~\bibnamefont
  {Jarrell}}\ and\ \bibinfo {author} {\bibfnamefont {J.}~\bibnamefont
  {Gubernatis}},\ }\href {\doibase
  http://dx.doi.org/10.1016/0370-1573(95)00074-7} {\bibfield  {journal}
  {\bibinfo  {journal} {Physics Reports}\ }\textbf {\bibinfo {volume} {269}},\
  \bibinfo {pages} {133 } (\bibinfo {year} {1996})}\BibitemShut {NoStop}%
\bibitem [{\citenamefont {Tiwari}\ and\ \citenamefont
  {Majumdar}(2013{\natexlab{a}})}]{triangular}%
  \BibitemOpen
  \bibfield  {author} {\bibinfo {author} {\bibfnamefont {R.}~\bibnamefont
  {Tiwari}}\ and\ \bibinfo {author} {\bibfnamefont {P.}~\bibnamefont
  {Majumdar}},\ }\href@noop {} {\bibfield  {journal} {\bibinfo  {journal}
  {arXiv preprint arXiv:1301.5026}\ } (\bibinfo {year}
  {2013}{\natexlab{a}})}\BibitemShut {NoStop}%
\bibitem [{\citenamefont {Tiwari}\ and\ \citenamefont
  {Majumdar}(2013{\natexlab{b}})}]{fcc}%
  \BibitemOpen
  \bibfield  {author} {\bibinfo {author} {\bibfnamefont {R.}~\bibnamefont
  {Tiwari}}\ and\ \bibinfo {author} {\bibfnamefont {P.}~\bibnamefont
  {Majumdar}},\ }\href@noop {} {\bibfield  {journal} {\bibinfo  {journal}
  {arXiv preprint arXiv:1302.2922}\ } (\bibinfo {year}
  {2013}{\natexlab{b}})}\BibitemShut {NoStop}%
\bibitem [{\citenamefont {Tarat}\ and\ \citenamefont
  {Majumdar}(2014)}]{tarat1}%
  \BibitemOpen
  \bibfield  {author} {\bibinfo {author} {\bibfnamefont {S.}~\bibnamefont
  {Tarat}}\ and\ \bibinfo {author} {\bibfnamefont {P.}~\bibnamefont
  {Majumdar}},\ }\href {http://stacks.iop.org/0295-5075/105/i=6/a=67002}
  {\bibfield  {journal} {\bibinfo  {journal} {EPL (Europhysics Letters)}\
  }\textbf {\bibinfo {volume} {105}},\ \bibinfo {pages} {67002} (\bibinfo
  {year} {2014})}\BibitemShut {NoStop}%
\bibitem [{\citenamefont {{Tarat}}\ and\ \citenamefont
  {{Majumdar}}(2014{\natexlab{a}})}]{tarat2}%
  \BibitemOpen
  \bibfield  {author} {\bibinfo {author} {\bibfnamefont {S.}~\bibnamefont
  {{Tarat}}}\ and\ \bibinfo {author} {\bibfnamefont {P.}~\bibnamefont
  {{Majumdar}}},\ }\href@noop {} {\bibfield  {journal} {\bibinfo  {journal}
  {arXiv preprints}\ } (\bibinfo {year} {2014}{\natexlab{a}})},\ \Eprint
  {http://arxiv.org/abs/1402.0817} {arXiv:1402.0817 [cond-mat.str-el]}
  \BibitemShut {NoStop}%
\bibitem [{\citenamefont {{Tarat}}\ and\ \citenamefont
  {{Majumdar}}(2014{\natexlab{b}})}]{tarat3}%
  \BibitemOpen
  \bibfield  {author} {\bibinfo {author} {\bibfnamefont {S.}~\bibnamefont
  {{Tarat}}}\ and\ \bibinfo {author} {\bibfnamefont {P.}~\bibnamefont
  {{Majumdar}}},\ }\href@noop {} {\bibfield  {journal} {\bibinfo  {journal}
  {arXiv preprints}\ } (\bibinfo {year} {2014}{\natexlab{b}})},\ \Eprint
  {http://arxiv.org/abs/1406.5423} {arXiv:1406.5423 [cond-mat.str-el]}
  \BibitemShut {NoStop}%
\bibitem [{\citenamefont {Schulz}(1990)}]{method-4}%
  \BibitemOpen
  \bibfield  {author} {\bibinfo {author} {\bibfnamefont {H.~J.}\ \bibnamefont
  {Schulz}},\ }\href {\doibase 10.1103/PhysRevLett.65.2462} {\bibfield
  {journal} {\bibinfo  {journal} {Phys. Rev. Lett.}\ }\textbf {\bibinfo
  {volume} {65}},\ \bibinfo {pages} {2462} (\bibinfo {year}
  {1990})}\BibitemShut {NoStop}%
\bibitem [{\citenamefont {Santos}(2003)}]{santos}%
  \BibitemOpen
  \bibfield  {author} {\bibinfo {author} {\bibfnamefont {R.~R.~d.}\
  \bibnamefont {Santos}},\ }\href
  {http://www.scielo.br/scielo.php?script=sci_arttext&pid=S0103-97332003000100003&nrm=iso}
  {\bibfield  {journal} {\bibinfo  {journal} {{Brazilian Journal of Physics}}\
  }\textbf {\bibinfo {volume} {33}},\ \bibinfo {pages} {36 } (\bibinfo {year}
  {2003})}\BibitemShut {NoStop}%
\bibitem [{\citenamefont {Staudt}\ \emph {et~al.}(2000)\citenamefont {Staudt},
  \citenamefont {Dzierzawa},\ and\ \citenamefont {Muramatsu}}]{muramatsu-1}%
  \BibitemOpen
  \bibfield  {author} {\bibinfo {author} {\bibfnamefont {R.}~\bibnamefont
  {Staudt}}, \bibinfo {author} {\bibfnamefont {M.}~\bibnamefont {Dzierzawa}}, \
  and\ \bibinfo {author} {\bibfnamefont {A.}~\bibnamefont {Muramatsu}},\ }\href
  {\doibase 10.1007/s100510070120} {\bibfield  {journal} {\bibinfo  {journal}
  {The European Physical Journal B - Condensed Matter and Complex Systems}\
  }\textbf {\bibinfo {volume} {17}},\ \bibinfo {pages} {411} (\bibinfo {year}
  {2000})}\BibitemShut {NoStop}%
\bibitem [{\citenamefont {Paiva}\ \emph {et~al.}(2001)\citenamefont {Paiva},
  \citenamefont {Scalettar}, \citenamefont {Huscroft},\ and\ \citenamefont
  {McMahan}}]{scalettar-1}%
  \BibitemOpen
  \bibfield  {author} {\bibinfo {author} {\bibfnamefont {T.}~\bibnamefont
  {Paiva}}, \bibinfo {author} {\bibfnamefont {R.~T.}\ \bibnamefont
  {Scalettar}}, \bibinfo {author} {\bibfnamefont {C.}~\bibnamefont {Huscroft}},
  \ and\ \bibinfo {author} {\bibfnamefont {A.~K.}\ \bibnamefont {McMahan}},\
  }\href {\doibase 10.1103/PhysRevB.63.125116} {\bibfield  {journal} {\bibinfo
  {journal} {Phys. Rev. B}\ }\textbf {\bibinfo {volume} {63}},\ \bibinfo
  {pages} {125116} (\bibinfo {year} {2001})}\BibitemShut {NoStop}%
\bibitem [{\citenamefont {Tahvildar-Zadeh}\ \emph {et~al.}(1997)\citenamefont
  {Tahvildar-Zadeh}, \citenamefont {Freericks},\ and\ \citenamefont
  {Jarrell}}]{wc2}%
  \BibitemOpen
  \bibfield  {author} {\bibinfo {author} {\bibfnamefont {A.~N.}\ \bibnamefont
  {Tahvildar-Zadeh}}, \bibinfo {author} {\bibfnamefont {J.~K.}\ \bibnamefont
  {Freericks}}, \ and\ \bibinfo {author} {\bibfnamefont {M.}~\bibnamefont
  {Jarrell}},\ }\href {\doibase 10.1103/PhysRevB.55.942} {\bibfield  {journal}
  {\bibinfo  {journal} {Phys. Rev. B}\ }\textbf {\bibinfo {volume} {55}},\
  \bibinfo {pages} {942} (\bibinfo {year} {1997})}\BibitemShut {NoStop}%
\bibitem [{\citenamefont {Scalettar}\ \emph {et~al.}(1989)\citenamefont
  {Scalettar}, \citenamefont {Scalapino}, \citenamefont {Sugar},\ and\
  \citenamefont {Toussaint}}]{wc1}%
  \BibitemOpen
  \bibfield  {author} {\bibinfo {author} {\bibfnamefont {R.~T.}\ \bibnamefont
  {Scalettar}}, \bibinfo {author} {\bibfnamefont {D.~J.}\ \bibnamefont
  {Scalapino}}, \bibinfo {author} {\bibfnamefont {R.~L.}\ \bibnamefont
  {Sugar}}, \ and\ \bibinfo {author} {\bibfnamefont {D.}~\bibnamefont
  {Toussaint}},\ }\href {\doibase 10.1103/PhysRevB.39.4711} {\bibfield
  {journal} {\bibinfo  {journal} {Phys. Rev. B}\ }\textbf {\bibinfo {volume}
  {39}},\ \bibinfo {pages} {4711} (\bibinfo {year} {1989})}\BibitemShut
  {NoStop}%
\bibitem [{\citenamefont {Georges}\ and\ \citenamefont
  {Krauth}(1993)}]{dmft-1}%
  \BibitemOpen
  \bibfield  {author} {\bibinfo {author} {\bibfnamefont {A.}~\bibnamefont
  {Georges}}\ and\ \bibinfo {author} {\bibfnamefont {W.}~\bibnamefont
  {Krauth}},\ }\href {\doibase 10.1103/PhysRevB.48.7167} {\bibfield  {journal}
  {\bibinfo  {journal} {Phys. Rev. B}\ }\textbf {\bibinfo {volume} {48}},\
  \bibinfo {pages} {7167} (\bibinfo {year} {1993})}\BibitemShut {NoStop}%
\bibitem [{\citenamefont {Chandra}\ \emph {et~al.}(1999)\citenamefont
  {Chandra}, \citenamefont {Kollar},\ and\ \citenamefont {Vollhardt}}]{dmft-2}%
  \BibitemOpen
  \bibfield  {author} {\bibinfo {author} {\bibfnamefont {N.}~\bibnamefont
  {Chandra}}, \bibinfo {author} {\bibfnamefont {M.}~\bibnamefont {Kollar}}, \
  and\ \bibinfo {author} {\bibfnamefont {D.}~\bibnamefont {Vollhardt}},\ }\href
  {\doibase 10.1103/PhysRevB.59.10541} {\bibfield  {journal} {\bibinfo
  {journal} {Phys. Rev. B}\ }\textbf {\bibinfo {volume} {59}},\ \bibinfo
  {pages} {10541} (\bibinfo {year} {1999})}\BibitemShut {NoStop}%
\bibitem [{\citenamefont {Vollhardt}(1997)}]{dmft-3}%
  \BibitemOpen
  \bibfield  {author} {\bibinfo {author} {\bibfnamefont {D.}~\bibnamefont
  {Vollhardt}},\ }\href {\doibase 10.1103/PhysRevLett.78.1307} {\bibfield
  {journal} {\bibinfo  {journal} {Phys. Rev. Lett.}\ }\textbf {\bibinfo
  {volume} {78}},\ \bibinfo {pages} {1307} (\bibinfo {year}
  {1997})}\BibitemShut {NoStop}%
\bibitem [{\citenamefont {Shiba}\ and\ \citenamefont {Pincus}(1972)}]{ed}%
  \BibitemOpen
  \bibfield  {author} {\bibinfo {author} {\bibfnamefont {H.}~\bibnamefont
  {Shiba}}\ and\ \bibinfo {author} {\bibfnamefont {P.~A.}\ \bibnamefont
  {Pincus}},\ }\href {\doibase 10.1103/PhysRevB.5.1966} {\bibfield  {journal}
  {\bibinfo  {journal} {Phys. Rev. B}\ }\textbf {\bibinfo {volume} {5}},\
  \bibinfo {pages} {1966} (\bibinfo {year} {1972})}\BibitemShut {NoStop}%
\bibitem [{\citenamefont {Duffy}\ and\ \citenamefont {Moreo}(1997)}]{moreo}%
  \BibitemOpen
  \bibfield  {author} {\bibinfo {author} {\bibfnamefont {D.}~\bibnamefont
  {Duffy}}\ and\ \bibinfo {author} {\bibfnamefont {A.}~\bibnamefont {Moreo}},\
  }\href {\doibase 10.1103/PhysRevB.55.12918} {\bibfield  {journal} {\bibinfo
  {journal} {Phys. Rev. B}\ }\textbf {\bibinfo {volume} {55}},\ \bibinfo
  {pages} {12918} (\bibinfo {year} {1997})}\BibitemShut {NoStop}%
\bibitem [{\citenamefont {Kumar}\ and\ \citenamefont
  {Majumdar}(2006{\natexlab{b}})}]{method-5}%
  \BibitemOpen
  \bibfield  {author} {\bibinfo {author} {\bibfnamefont {S.}~\bibnamefont
  {Kumar}}\ and\ \bibinfo {author} {\bibfnamefont {P.}~\bibnamefont
  {Majumdar}},\ }\href {\doibase 10.1140/epjb/e2006-00173-2} {\bibfield
  {journal} {\bibinfo  {journal} {The European Physical Journal B - Condensed
  Matter and Complex Systems}\ }\textbf {\bibinfo {volume} {50}},\ \bibinfo
  {pages} {571} (\bibinfo {year} {2006}{\natexlab{b}})}\BibitemShut {NoStop}%
\bibitem [{\citenamefont {Duffy}\ \emph {et~al.}(1997)\citenamefont {Duffy},
  \citenamefont {Nazarenko}, \citenamefont {Haas}, \citenamefont {Moreo},
  \citenamefont {Riera},\ and\ \citenamefont {Dagotto}}]{uttp-1}%
  \BibitemOpen
  \bibfield  {author} {\bibinfo {author} {\bibfnamefont {D.}~\bibnamefont
  {Duffy}}, \bibinfo {author} {\bibfnamefont {A.}~\bibnamefont {Nazarenko}},
  \bibinfo {author} {\bibfnamefont {S.}~\bibnamefont {Haas}}, \bibinfo {author}
  {\bibfnamefont {A.}~\bibnamefont {Moreo}}, \bibinfo {author} {\bibfnamefont
  {J.}~\bibnamefont {Riera}}, \ and\ \bibinfo {author} {\bibfnamefont
  {E.}~\bibnamefont {Dagotto}},\ }\href {\doibase 10.1103/PhysRevB.56.5597}
  {\bibfield  {journal} {\bibinfo  {journal} {Phys. Rev. B}\ }\textbf {\bibinfo
  {volume} {56}},\ \bibinfo {pages} {5597} (\bibinfo {year}
  {1997})}\BibitemShut {NoStop}%
\bibitem [{\citenamefont {Nazarenko}\ \emph {et~al.}(1995)\citenamefont
  {Nazarenko}, \citenamefont {Vos}, \citenamefont {Haas}, \citenamefont
  {Dagotto},\ and\ \citenamefont {Gooding}}]{uttp-2}%
  \BibitemOpen
  \bibfield  {author} {\bibinfo {author} {\bibfnamefont {A.}~\bibnamefont
  {Nazarenko}}, \bibinfo {author} {\bibfnamefont {K.~J.~E.}\ \bibnamefont
  {Vos}}, \bibinfo {author} {\bibfnamefont {S.}~\bibnamefont {Haas}}, \bibinfo
  {author} {\bibfnamefont {E.}~\bibnamefont {Dagotto}}, \ and\ \bibinfo
  {author} {\bibfnamefont {R.~J.}\ \bibnamefont {Gooding}},\ }\href {\doibase
  10.1103/PhysRevB.51.8676} {\bibfield  {journal} {\bibinfo  {journal} {Phys.
  Rev. B}\ }\textbf {\bibinfo {volume} {51}},\ \bibinfo {pages} {8676}
  (\bibinfo {year} {1995})}\BibitemShut {NoStop}%
\bibitem [{\citenamefont {Sentef}\ \emph {et~al.}(2011)\citenamefont {Sentef},
  \citenamefont {Werner}, \citenamefont {Gull},\ and\ \citenamefont
  {Kampf}}]{uttp-3}%
  \BibitemOpen
  \bibfield  {author} {\bibinfo {author} {\bibfnamefont {M.}~\bibnamefont
  {Sentef}}, \bibinfo {author} {\bibfnamefont {P.}~\bibnamefont {Werner}},
  \bibinfo {author} {\bibfnamefont {E.}~\bibnamefont {Gull}}, \ and\ \bibinfo
  {author} {\bibfnamefont {A.~P.}\ \bibnamefont {Kampf}},\ }\href {\doibase
  10.1103/PhysRevLett.107.126401} {\bibfield  {journal} {\bibinfo  {journal}
  {Phys. Rev. Lett.}\ }\textbf {\bibinfo {volume} {107}},\ \bibinfo {pages}
  {126401} (\bibinfo {year} {2011})}\BibitemShut {NoStop}%
\bibitem [{\citenamefont {Tocchio}\ \emph {et~al.}(2008)\citenamefont
  {Tocchio}, \citenamefont {Becca}, \citenamefont {Parola},\ and\ \citenamefont
  {Sorella}}]{gs-1}%
  \BibitemOpen
  \bibfield  {author} {\bibinfo {author} {\bibfnamefont {L.~F.}\ \bibnamefont
  {Tocchio}}, \bibinfo {author} {\bibfnamefont {F.}~\bibnamefont {Becca}},
  \bibinfo {author} {\bibfnamefont {A.}~\bibnamefont {Parola}}, \ and\ \bibinfo
  {author} {\bibfnamefont {S.}~\bibnamefont {Sorella}},\ }\href {\doibase
  10.1103/PhysRevB.78.041101} {\bibfield  {journal} {\bibinfo  {journal} {Phys.
  Rev. B}\ }\textbf {\bibinfo {volume} {78}},\ \bibinfo {pages} {041101}
  (\bibinfo {year} {2008})}\BibitemShut {NoStop}%
\bibitem [{\citenamefont {Becca}\ \emph {et~al.}(2009)\citenamefont {Becca},
  \citenamefont {Tocchio},\ and\ \citenamefont {Sorella}}]{gs-2}%
  \BibitemOpen
  \bibfield  {author} {\bibinfo {author} {\bibfnamefont {F.}~\bibnamefont
  {Becca}}, \bibinfo {author} {\bibfnamefont {L.~F.}\ \bibnamefont {Tocchio}},
  \ and\ \bibinfo {author} {\bibfnamefont {S.}~\bibnamefont {Sorella}},\ }\href
  {http://stacks.iop.org/1742-6596/145/i=1/a=012016} {\bibfield  {journal}
  {\bibinfo  {journal} {Journal of Physics: Conference Series}\ }\textbf
  {\bibinfo {volume} {145}},\ \bibinfo {pages} {012016} (\bibinfo {year}
  {2009})}\BibitemShut {NoStop}%
\bibitem [{\citenamefont {Markiewicz}\ \emph {et~al.}(2010)\citenamefont
  {Markiewicz}, \citenamefont {Lorenzana},\ and\ \citenamefont
  {Seibold}}]{gs-3}%
  \BibitemOpen
  \bibfield  {author} {\bibinfo {author} {\bibfnamefont {R.~S.}\ \bibnamefont
  {Markiewicz}}, \bibinfo {author} {\bibfnamefont {J.}~\bibnamefont
  {Lorenzana}}, \ and\ \bibinfo {author} {\bibfnamefont {G.}~\bibnamefont
  {Seibold}},\ }\href {\doibase 10.1103/PhysRevB.81.014510} {\bibfield
  {journal} {\bibinfo  {journal} {Phys. Rev. B}\ }\textbf {\bibinfo {volume}
  {81}},\ \bibinfo {pages} {014510} (\bibinfo {year} {2010})}\BibitemShut
  {NoStop}%
\end{thebibliography}%
\end{document}